\documentclass[pdflatex,sn-vancouver-num,Numbered,referee]{sn-jnl}

\usepackage{graphicx}
\usepackage{multirow}
\usepackage{amsmath,amssymb,amsfonts}
\usepackage{amsthm}
\usepackage[title]{appendix}
\usepackage{xcolor}
\usepackage{textcomp}
\usepackage{manyfoot}
\usepackage{booktabs}
\usepackage{algorithm}
\usepackage{algorithmicx}
\usepackage{algpseudocode}
\usepackage{listings}
\usepackage{siunitx}
\usepackage{mathtools}
\usepackage{subcaption}
\usepackage[section]{placeins}

\theoremstyle{thmstyleone}

\theoremstyle{thmstyletwo}

\theoremstyle{thmstylethree}

\begin{document}

\title[Slow dynamics in concrete: Effects of temperature, strength variation, and microcracking damage]{Slow dynamics in concrete: Effects of temperature, strength variation, and microcracking damage}

\author[1]{\fnm{Clayton} \sur{Malone}}\email{cmalone@huskers.unl.edu}

\author*[1]{\fnm{Jinying} \sur{Zhu}}\email{jyzhu@unl.edu}

\affil*[1]{\orgdiv{Department of Civil and Environmental Engineering}, \orgname{University of Nebraska-Lincoln}, \orgaddress{\street{1110 S 67TH ST}, \city{Omaha}, \postcode{68182}, \state{NE}, \country{USA}}}

\abstract{
In this study, we investigated the slow dynamic behavior of concrete with varying compressive strengths ($f'_c = 32\sim56$ MPa) and alkali-silica reaction (ASR) damage.
Concrete prisms were conditioned by compressive loading and recovery was monitored by coda wave interferometry (CWI). 
A self-referencing temperature correction technique was used to minimize the effect of ambient temperature changes, demonstrating the importance of careful temperature control and correction.
For intact specimens, both the recovery rate, $m_v$, and velocity drop magnitude, $|c|$, generally increased with compressive strength.
Recovery times were calculated from the fit parameters, and higher strength specimens were found to recover faster (9.9 h for $f'_c = 32$ MPa, 6.6 h for $f'_c = 56$ MPa).
When the same stress was applied, ASR-damaged specimens had increased softening and faster recovery rates, but longer recovery times (up to 458 h).
These findings demonstrate the potential of slow dynamics in concrete characterization, through its ability to evaluate the strength of intact samples and its sensitivity to microcracking in damaged samples.
}


\maketitle
\thanks{\small\textit{{This manuscript is under review at the Journal of Nondestructive Evaluation.}}

\section{Introduction}\label{sec:1}
Concrete is widely used in critical infrastructure such as bridges, dams, and nuclear power plants, but deteriorates over time, leading to reduced structural performance and safety.
The compressive strength of concrete is a key indicator of structural performance~\cite{pucinotti2013assessment}, while deterioration mechanisms such as alkali-silica reaction (ASR) cause a gradual degradation of material properties~\cite{ahmed2003effect}.
Accurate assessment of material properties is essential for maintenance planning and service life prediction, particularly in safety-critical infrastructure like nuclear-containment structures~\cite{alexander2019durability}.
Direct assessment of concrete through core sampling enables accurate characterization but is invasive, costly, and often impractical for large or heavily reinforced members.
These limitations have led to the widespread development of nondestructive testing (NDT) methods, which offer more time- and resource-efficient means of assessing damage and estimating remaining service life~\cite{alqurashi2025review}.

Conventional NDT methods, such as ultrasonic pulse velocity, rebound hammer, or impact-echo, can be used to estimate the stiffness of concrete.
However, reliance on linear wave propagation limits their ability to detect subtle microstructure changes.
Additionally, their results often depend on empirical correlations that require parameters that may not be known, such as mix properties, age, or environmental conditions~\cite{trtnik2009prediction}.
This leads to estimates of material properties that are localized and affected by field conditions.
To improve the accuracy of conventional methods, machine learning models have been developed using large, aggregated databases that incorporate extensive field data~\cite{matthews2026advancing}.
However, current models continue to face challenges in accurately predicting behavior when conditions and mix designs fall outside their training domain~\cite{taffese2025machine}.

Nonlinear acoustic (NLA) methods measure strain dependent responses and are especially sensitive to microstructure changes in heterogeneous materials~\cite{guyer1999nonlinear}.
These methods have been used to characterize damage in concrete, with sensitivity that can be an order of magnitude higher than linear methods~\cite{shkolnik_2005, payan2007applying, chen2010rapid, kim2018situ, malone2021evaluation}. 
Concrete’s strength and durability are governed by its complex microstructure, composed of cracks, cement paste, aggregates, the interfacial transition zone (ITZ), and pore network~\cite{das2012implication, vandenabeele2002influence}.
Deterioration mechanisms like ASR create expansive reaction products that result in cracks and porosity changes, further complicating the microstructure.
Linear ultrasonic parameters (such as wave velocities, resonance frequencies) are only weakly sensitive to these features, unlike nonlinear approaches, which evaluate changes in those linear parameters under varying strain levels~\cite{johnson1996resonance,vandenabeele2000nonlinear_1, vandenabeele2000nonlinear_2}. 
Shkolnik~\cite{shkolnik_2005} provided a physical basis for this enhanced sensitivity, demonstrating that nonlinear parameters are markedly more responsive to microstructural features such as microcrevices and micropores than linear wave velocities.
Furthermore, nonlinear parameters were found to be two to three orders of magnitude larger in concrete than in crystalline or polycrystalline materials.
Common NLA approaches include nonlinear resonant ultrasound spectroscopy (NRUS)~\cite{vandenabeele2000nonlinear_1, vandenabeele2000nonlinear_2}, nonlinear impact resonance acoustic spectroscopy (NIRAS)~\cite{chen2010rapid, lesnicki2011characterization}, and dynamic acousto-elastic testing (DAET)~\cite{riviere2013pump}.
These techniques assess  the material's nonlinear response during or immediately following an applied perturbation.
In contrast, the phenomenon of slow dynamics describes the subsequent recovery process after a perturbation ceases~\cite{tencate1996slow, tencate2011slow}.

To understand the significance of slow dynamics in concrete, it is helpful to review its origins in geomaterials~\cite{guyer1999nonlinear, tencate1996slow, tencate2011slow}.
The phenomenon was first reported in resonance experiments on Berea sandstone in the mid 1990s, described as a form of material memory where the elastic state depends on its recent strain history~\cite{guyer1995hysteresis, tencate1996slow}.
It involves two distinct phases: conditioning, where a dynamic strain results in a rapid drop in stiffness (fast dynamics), and relaxation, where the stiffness slowly recovers (slow dynamics) after the strain is removed~\cite{tencate1996slow, tencate2000slow}.
Although first explored in the field of geophysics~\cite{tencate2011slow}, the study of this behavior has been extended to a wide range of materials, including concrete~\cite{bentahar2006hysteretic, kodjo2011impact, tremblay2010probing,larose2013ultrasonic}, ceramics~\cite{guyer1999nonlinear, johnson2005slow}, glass~\cite{yoritomo2020slow_1, yoritomo2020slow_2, bittner2022transient}, and metals~\cite{johnson2005slow, yoritomo2020slow_3, kober2022material}.

The physical origin of slow dynamics is attributed to the collective response of the material's ``bond system".
The bond system is a network of soft mesoscopic features, such as intergrain contacts and microcracks, within the hard material matrix~\cite{guyer1999nonlinear, johnson2005slow, guyer2009nonlinear}.
The healing of this network of features occurs at different rates, causing the observed log(time) recovery of the material's modulus~\cite{tencate2000universal, snieder2017time}.
However, the exact physical mechanisms at play remain a subject of investigation.
Recent work models this multi-rate process using a continuous relaxation-time spectrum, showing its peak is determined by material grain size and is independent of conditioning amplitude~\cite{kober2022material}.
Others have focused on the role of moisture, providing direct visual evidence of microscale moisture migration during recovery~\cite{bittner2019direct}.
The Mechanistic Diffusion Model (MDM) describes this behavior, proposing that it is driven by diffusion of moisture vapor near grain contacts~\cite{bittner2021mechanistic}.
Further proposed mechanisms include thermally activated healing of bonds and friction at crack surfaces~\cite{tencate2000slow}.
Recent experimental evidence supports a mechanism in which slow dynamics is attributed to frictional processes at oblique sliding contacts, with the subsequent recovery related to contact aging rather than the opening and closing of cracks~\cite{asnar2025anisotropy}.

The study of the mechanisms that drive slow dynamics has been accompanied by the development of experimental approaches for measuring its response.
Techniques include NRUS to track amplitude dependent frequency shifts~\cite{johnson1996resonance, tencate2000slow, johnson2005slow, bentahar2006hysteretic, kober2022material}, and pump-probe tests such as DAET that measure the material's hysteretic stress-strain behavior~\cite{riviere2013pump, shokouhi2017dynamic, shokouhi2017slow}.
In high scattering materials, diffuse ultrasound has been applied using coda wave interferometry (CWI) to measure relative velocity changes~\cite{tremblay2010probing, larose2013ultrasonic, shokouhi2017slow, yoritomo2020slow_1, yoritomo2020slow_2} or diffuse acoustic wave spectroscopy (DAWS) to monitor rearrangement of grain contacts~\cite{bissig2003intermittent, tremblay2010probing, larose2013ultrasonic}.
Other methods include sequential impact tests~\cite{bittner2018understanding,bittner2022transient}, and techniques like Larsen effect monitoring~\cite{lobkis2009larsen} and the digital lock-in probe~\cite{yoritomo2025slow} which enable the study of recovery at early times.

The sensitivity of slow dynamics to microstructural features has motivated its application to concrete, a heterogeneous material for which conventional characterization methods often have limited sensitivity.
Experimental studies have shown that concrete displays slow dynamic behavior under low-level acoustic~\cite{johnson2005slow, bentahar2006hysteretic} and mechanical perturbations~\cite{tremblay2010probing, larose2013ultrasonic}, while thermal perturbations have been shown to induce slow dynamics in concrete~\cite{tencate2000universal} and Berea sandstone~\cite{simpson2023temperature}, with these responses associated with thermally induced strain from differential thermal expansion/contraction.
In concrete, this response is sensitive to microstructure changes caused by mechanical damage~\cite{bentahar2006hysteretic, tremblay2010probing}, ASR~\cite{kodjo2011impact}, and thermal damage~\cite{larose2013ultrasonic, bekele2017slow, bittner2022transient}.
While most slow dynamics studies on concrete have focused on damage characterization, fewer have explored its relation to material properties such as compressive strength. 
Significant research has already established links between slow dynamics parameters and other specific material states, such as the presence of ASR gel~\cite{kodjo2011impact}, moisture state and diffusion processes~\cite{bittner2019direct,bittner2021mechanistic}, and changing elastic properties during cement paste curing~\cite{lobkis2009larsen}. In addition, DAET has been used to estimate classical nonlinear elastic parameters ($\beta$, $\delta$) in intact concrete~\cite{shokouhi2017dynamic}, demonstrating the potential of probing material properties. Despite these advances, few studies have examined whether slow dynamics parameters are related to concrete strength.  Building on the established theoretical link between compressive strength and the nonlinear stress-strain response~\cite{shkolnik_2005}, this study investigates whether the slow dynamics recovery parameters are similarly governed by the material's strength and microstructure.

Motivated by these developments, this study evaluates slow dynamics as an NDT technique for characterizing concrete specimens with varied strengths and ASR-induced damage.
Accordingly, we monitored the recovery behavior of intact and damaged specimens and related the measured parameters to specimen strengths and damage levels.
Section~\ref{sec:2} introduces the theoretical framework, including the recovery model, reference time estimation, CWI analysis, and temperature correction procedure.
Section~\ref{sec:3} details the specimens and Section~\ref{sec:4} gives the experimental setups.
Section~\ref{sec:5} presents the results of temperature correction, slow dynamics characterization across mixes of varying strengths, and responses of ASR-damaged specimens.
Section~\ref{sec:6} summarizes the key findings, stating implications for strength estimation and damage detection, and offers recommendations for practical applications.
\section{Theoretical background and signal analysis methods}\label{sec:2}

\subsection{Slow dynamics}\label{sec:2.1}
This study examines stress-induced slow dynamics, where applied stress causes a modulus reduction that is followed by logarithmic recovery upon removal.
This recovery process can be described generally as:
\begin{equation}
\label{equation 1}
\frac{\Delta E}{E} = m_E \cdot \log(t) + c_E
\end{equation}
where $E$ is the material modulus, $t$ is the elapsed time since the stress was removed, and $m_E$ is a constant describing the modulus recovery rate (slope).
Here and throughout this work, the time $t$ was normalized by the unit time 1 second.
Therefore, the logarithmic term is dimensionless, and the intercept parameter $c_E$ represents the extrapolated relative modulus change at 1 second after the end of conditioning.

However, because the modulus itself is difficult to directly measure, resonance frequency or relative velocity change ($dV/V$) is typically used to represent the modulus change.
Taking relative velocity change as a probe for the modulus, the equation can be written as:
\begin{equation}
\label{equation 2}
\frac{dV}{V} = m_v \cdot \log(t) + c
\end{equation}
where $m_v$ is a constant used to describe the recovery rate of $dV/V$, and $c$ represents the extrapolated relative velocity change value at 1 second after the end of conditioning.
Because $c$ is negative following conditioning, its magnitude, $|c|$, is used when comparing the velocity drop between specimens.

For convenience, a recovery time $t_r$ can be defined as the time at which the fitted $dV/V$ returns to zero, i.e., the implied time needed for the material to recover to its initial, preconditioned state~\cite{tremblay2010probing}.
Setting Eq.~\ref{equation 2} to zero gives:
\begin{equation}
\label{equation 3}
t_r = 10^{(-c/m_v)} 
\end{equation}

Fig.~\ref{fig:Intro} provides an example of monitoring the recovery of a concrete specimen.
When a compressive load is applied, $dV/V$ increases rapidly due to the acoustoelastic effect.
As the load is removed, $dV/V$ drops sharply due to both the acoustoelastic effect and conditioning during loading, falling below the preconditioned baseline.
A slow log(time) recovery follows.

\begin{figure}[!htb]        
\centering 
\includegraphics[width=3in]{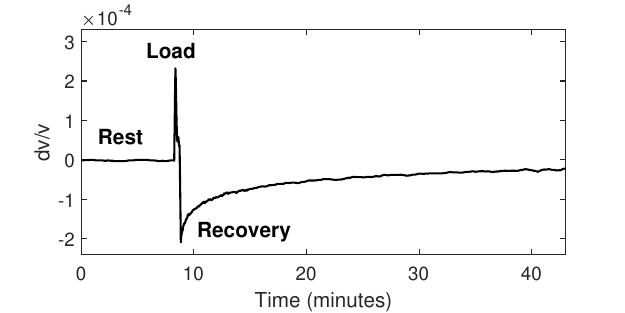}        
\caption{Example of relative velocity change ($dV/V$) in a concrete specimen during rest, loading, and recovery phases of a slow dynamics test.}
\label{fig:Intro} 
\end{figure}
\FloatBarrier
\subsection{Reference time estimation}\label{sec:2.2}
In slow dynamics behavior, the modulus drop and subsequent recovery begin when the applied stress is removed from the material.
The time when conditioning ends and recovery begins, known as the reference time, is denoted by $t_0$.
Lee and Weaver~\cite{lee2024slow} performed a study detailing the importance of $t_0$ selection and showed that the linear log(time) relationship can be maintained to early times in the range of milliseconds when $t_0$ is appropriately selected.
They described the challenges associated with $t_0$ selection, noting that measurements assumed to represent recovery may still be contaminated by residual effects from conditioning.
A further complexity arises during ringdown where conditioning and recovery occur simultaneously.
This challenge was addressed in a recent study by Yoritomo and Weaver~\cite{yoritomo2025slow} where they developed a model that describes the modulus evolution during conditioning, ringdown, and recovery.
In summary, the choice of $t_0$ is critical and readers are directed to the aforementioned studies for detailed discussions and recommended procedures.

This study implemented the procedure of Sun~\cite{sun2020thesis} by selecting $t_0$ to be that which maximizes the linearity of recovery with respect to log(time).
The search was performed starting from the point of full load removal and iterating over possible $t_0$ values, selecting the one that provides the best log(time) fit within the first 30 s of recovery.
Measurements beyond 30 s were excluded in $t_0$ selection to minimize influence from external factors, such as temperature variations, that can have a significant influence across longer time scales.
A trade-off between the number of measurements included in the fit and the influence of external factors exists, and a shorter or longer time range may be suitable in different settings.
Further discussion of temperature influence is found in Section~\ref{sec:2.3}.

Once estimated, the time axis can be adjusted so that the first recorded measurement during recovery begins at a time of $t-t_0$. To account for this, Eq.~\ref{equation 2} is modified:
\begin{equation}
\label{equation 4}
\frac{dV}{V} = m_v \cdot \log(t-t_0) + c
\end{equation}
The procedure used for selecting $t_0$ can be seen in Fig.~\ref{fig:Conditioning and recovery}(a). Once conditioning is known to have ceased, the first point of recovery was taken and its time was offset by $t_0$. Selecting an excessively early or late $t_0$ value will lead to errors at short times, as shown in Fig.~\ref{fig:Conditioning and recovery}(b).

\begin{figure}[!htb]
\centering
\begin{subfigure}{0.56\textwidth}
    \centering
    \includegraphics[width=2.66in]{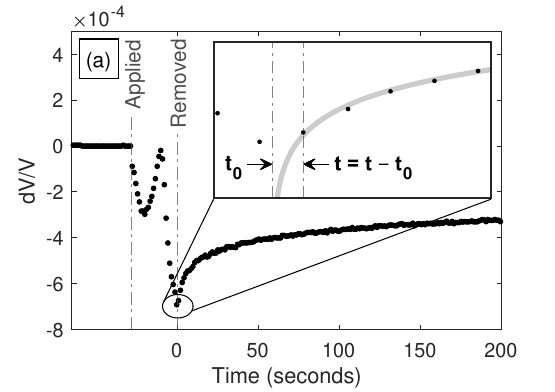} 
\end{subfigure}
\begin{subfigure}{0.44\textwidth}
    \centering
    \includegraphics[width=2.04in]{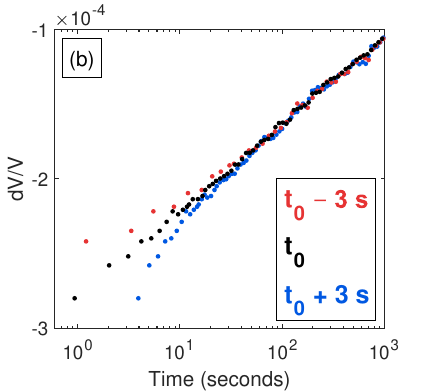}
\end{subfigure}
\caption{Example of recovery reference time estimation. (a) Example of $t_0$ selection following load removal. (b) Effect of selecting $t_0$ too early or late on log(time) recovery.}
\label{fig:Conditioning and recovery} 
\end{figure}
\FloatBarrier
\subsection{Influence of temperature and self-temperature correction}\label{sec:2.3}
Prior studies have described the effect that temperature fluctuations can have on slow dynamics experimental results.
In an early work, TenCate et al.~\cite{tencate2000universal} noted slow dynamics could be induced not only by mechanical strain but also by a sudden temperature change, which was also observed by Sun and Zhu~\cite{sun2019thermal}.
Subsequent works documented the effects of temperature and humidity variations on rock recovery~\cite{tencate2002nonlinearity} and emphasized the need for careful environmental control, performing experiments under passive thermal isolation~\cite{tencate2004nonlinear}.
Likewise, Johnson and Sutin~\cite{johnson2005slow} reiterated the importance of thermal control, noting that extreme care is needed in controlling the effect of temperature in slow dynamics tests.  

The extent to which thermal effects influence experimental results depends on the material being tested and the magnitude of temperature variation during testing.
Recovery can be isolated from these effects by maintaining a stable thermal environment, but this is often impractical outside controlled laboratory settings.
Even small temperature changes (on the order of 0.1$^\circ$C for concrete) can influence results, demonstrating the need for correction procedures when thermal control is not feasible.

Studies by Sun and Zhu~\cite{sun2019thermal,sun2020determination} estimated the acoustic nonlinearity parameters of thermal- and ASR-damaged concrete specimens by measuring the relationship between wave velocity and temperature, finding that damaged specimens were more sensitive to temperature changes than undamaged specimens.
If the temperature sensitivity of $dV/V$ of concrete is known and the temperature history is measured, its response can be corrected for thermal effects.
Building on this concept, we applied a temperature correction procedure that we previously developed for acoustoelastic experiments on metal specimens~\cite{zeng2023temperature}.
The procedure is self-referencing, meaning only the specimen being tested is needed in the correction process.
This offers an advantage over methods that require a reference specimen, since reference samples do not experience the same temperature effects as the test specimen and are likely unavailable in field testing. 

Acknowledging that the effects of mechanical and thermal changes are decoupled, i.e., they act independently, Eq.~\ref{equation 4} can be modified to show the measured relative velocity change is a ``mixed" value with contributions from both mechanically induced slow dynamics recovery and thermal effects:
\begin{equation}
\label{equation 5}
\underbracket{\left(\frac{dV^M}{V^0}\right)}_{\text{measured}}
=
\underbracket{\left(\vphantom{\frac{dV^T/V^0}{dT}} m_v \cdot\log(t - t_0) + c \right)}_{\text{recovery}}
+
\underbracket{\left(\frac{dV^T/V^0}{dT} \cdot \Delta T\right)}_{\text{thermal}}
\end{equation}
where $\Delta T$ represents the temperature change that occurred during recovery and $\frac{{{{d{V^T}} \mathord{\left/
{\vphantom {{d{V^T}} {{V^0}}}} \right.
\kern-\nulldelimiterspace} {{V^0}}}}}{dT}$ is the relative velocity change due to temperature, known as the thermo-acoustic coefficient (TAC).

The TAC of a concrete specimen can be obtained by monitoring the relative velocity change of the sample under the influence of only temperature changes, referred to as a thermal modulation test.
After obtaining the TAC, temperature correction for each measurement can be performed by subtracting the thermal contribution from the measured ``mixed" $dV/V$.
This yields the ``corrected" $dV/V$ that represents the contribution from mechanically induced slow dynamics behavior only.
\begin{equation}
\label{equation 6}
\frac{dV^R}{V^0} = \frac{dV^M}{V^0} - \frac{dV^T/V^0}{dT} \cdot \Delta T
\end{equation}
where the superscripts $R$, $M$, and $T$ denote the recovery (corrected), measured (mixed), and thermal contributions, respectively.

It should be noted that concrete, a nonlinear hysteretic material, has unique TACs for heating and cooling.
In this study both the ``heating" and ``cooling" TACs were determined and used for correction corresponding to the type of thermal change experienced.
Additionally, the temperature fluctuations experienced in this study were gradual, different from the sudden ``thermal shocking" that can independently induce softening and recovery.
For cases involving non-monotonic temperature changes or rapid changes that induce slow dynamics, a different correction procedure would likely be required.
\FloatBarrier
\subsection{Velocity change analysis by stretching technique}\label{sec:2.4} 
In this study, $dV/V$ was obtained using the stretching technique, a common approach in CWI analysis~\cite{lobkis2003codawave}, which assumes that the perturbed signal $s_1(t)$ is a time stretched or compressed version of the reference signal $s_0(t)$.
By calculating the cross-correlation coefficient (CC) between the two signals, the relative velocity change between them can be determined.
The process is expressed as:
\begin{equation}
\begin{aligned}
CC\left( \varepsilon  \right) = \frac{{\int_{{t_{a}}}^{{t_{b}}} {{s_1}\left[ {t\left( {1 + \varepsilon } \right)} \right]{s_0}\left( t \right)dt} }}{{\sqrt {\int_{{t_a}}^{{t_b}} {s_1^2\left[ {t\left( {1 + \varepsilon } \right)} \right]dt} \int_{{t_a}}^{{t_b}} {s_0^2\left( t \right)dt} } }}
\label{equation 7}
\end{aligned}
\end{equation}
where $[t_a, t_b]$ defines the chosen time window of the signal for the analysis, and $\varepsilon$ is the stretching factor.
The relative velocity change is calculated from the relative time change by $dV/V=-dt/t=\varepsilon_{max}$, when $\varepsilon_{max}$ maximizes the CC.
The stretching technique has been used to measure stress and temperature induced velocity changes in concrete~\cite{niederleithinger2013influence, niederleithinger2018processing}.
Although CWI was originally developed for analyzing coda waves, it has also been extended to coherent waves.
Its limitations and sources of error have been examined in detail in a recent publication~\cite{zeng2026calculation}. 
\FloatBarrier
\section{Materials}\label{sec:3}
\subsection{Intact specimens}\label{sec:3.1}
The four primary mix designs for the concrete prisms (75 mm × 75 mm × 285 mm) tested in this study are summarized in Table~\ref{tab:mix_designs}, along with batch properties.
The coarse aggregate (limestone) and fine aggregate (river sand) properties are given in Table~\ref{tab:aggregate_properties}.
All mixes were cast with Ash Grove Type I/II Portland cement.
The primary intact specimens were moist cured for 28 days and tested approximately 18 months after casting.
After moist curing, the specimens were stored together in a closed, unpowered chamber under ambient laboratory conditions (nominally 23$^\circ$C, with relative humidity typically ranging from 15\% to 50\%).
Relative humidity was not actively controlled during storage or testing.

For each mix design, slow dynamics tests were performed on three prisms.
Fig.~\ref{fig:Specimens} shows a prism from each mix.
Cylinders from each mix were tested for compressive strength in accordance with ASTM C39, yielding strengths of 32.2, 40.4, 48.8, and 56.3 MPa, respectively~\cite{ASTM-C39-24}.
\begin{table}[!htb]
\centering
\caption{Concrete mix designs (SSD) and selected properties.}
\begin{tabular}{lrrrr}
\toprule
Ingredient                  & Mix 1 & Mix 2 & Mix 3 & Mix 4 \\
\midrule
Cement (kg/m$^3$)           & $263$   & $341$   & $421$   & $534$  \\
Coarse Aggregate (kg/m$^3$) & $1100$  & $1100$  & $1100$  & $1100$ \\
Fine Aggregate (kg/m$^3$)   & $793$   & $727$   & $659$   & $565$  \\
Water (kg/m$^3$)            & $160$   & $160$   & $160$   & $160$  \\
HRWR (mL)                   & -       & -       & -       & $50$   \\
w/c                         & $0.61$  & $0.47$  & $0.38$  & $0.30$ \\
\midrule
Slump (mm)                  & $115$   & $120$   & $75$    & $150$ \\
Unit Weight (kg/m$^3$)      & $2413$  & $2384$  & $2459$  & $2428$ \\
\midrule
28-day $f'_{c}$ (MPa)       & $32.2$  & $40.4$  & $48.8$  & $56.3$ \\
$f'_{c}$ CV (\%)            & $3.8$   & $4.0$   & $2.6$   & $1.4$  \\
\bottomrule
\end{tabular}
\label{tab:mix_designs}
\end{table}
\begin{table}[!htb]
\centering
\caption{Properties of fine and coarse aggregates used in Mix 1-4 (SSD).} 
{\small
\begin{tabular}{lccc}
\hline
Type               & SG     & Absorption   & DRUW \\
                   & (SSD)  &              & (kg/m$^3$)\\
\hline
Fine aggregate     & $2.64$	& $0.42$\%     & - \\
Coarse aggregate   & $2.55$	& $2.57$\%     & $1660$ \\
\hline
\end{tabular}}
\label{tab:aggregate_properties}
\end{table} 
\begin{figure}[!htb]        
\centering 
\includegraphics[width=3in]{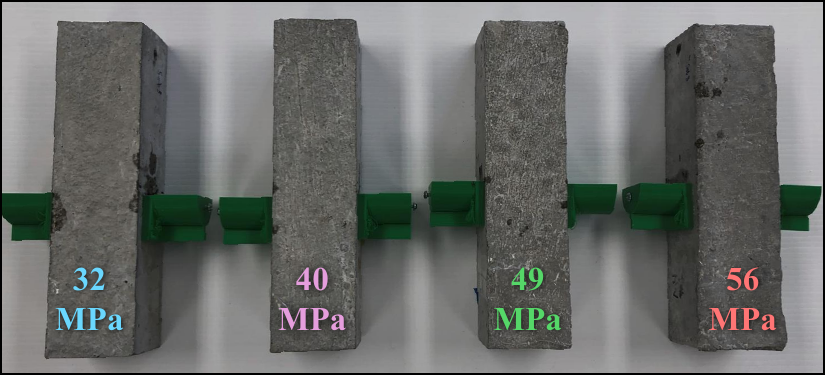}        
\caption{Concrete prisms used in the study, labeled by nominal compressive strength. Three prisms from each batch were tested.}
\label{fig:Specimens} 
\end{figure}
\FloatBarrier
\subsection{Additional intact specimens}\label{sec:3.2}
Two additional intact specimens were tested to evaluate whether the relationships between recovery parameters and compressive strength from the primary group of specimens (see Sec.~\ref{sec:5.2.1}) are broadly applicable across different specimen ages and mix designs.
Although the specific mix designs and aggregate properties are unknown, the specimens (referred to as Specimen \#1 and Specimen \#2) were cast approximately five years prior to the primary groups and had batch compressive strengths of 34.1 MPa and 35.0 MPa, respectively.
Before testing, these specimens were stored under the same ambient laboratory conditions as the primary intact specimens. Although relative humidity was not actively controlled during storage, the moisture content measured on companion specimens cast and stored in the same conditions was 2.5\% by mass.
\subsection{ASR-damaged specimens}\label{sec:3.3}
Two concrete specimens with ASR-induced damage (Fig.~\ref{fig:ASR_specimens}) were also tested.  
The specimens are identified by the type of reactive aggregate used in their mix: RFA refers to the specimen with a reactive fine aggregate and innocuous coarse aggregate, and RCA refers to the specimen with a reactive coarse aggregate and innocuous fine aggregate.
For both mix designs, NaOH was added to raise the alkali content to 1.25\% Na$_2$O$_{\mathrm{eq}}$ by mass of cement, i.e., 5.25 kg/m$^3$ Na$_2$O$_{\mathrm{eq}}$, promoting accelerated ASR.
Aggregates were prepared and graded in accordance with ASTM C1293.
Both specimens had batch compressive strengths of approximately 35 MPa.

The ASR-damaged specimens were cast approximately five years before the present testing and were investigated by the authors in a previous study~\cite{malone2021evaluation}.
At the conclusion of that study, the increased temperature/humidity conditioning was stopped, and the specimens were stored under the same ambient laboratory conditions as the intact specimens.
Relative humidity was not actively controlled after conditioning or during testing.
Petrographic analysis showed distinct damage mechanisms in the two specimens.
The RCA specimen had cracking throughout the coarse aggregate, forming a network of macrocracks, whereas the RFA specimen displayed microcracking primarily in the cement matrix and ITZ.
The expansion levels at the time of this study were 0.0721\% and 0.0230\% for the RCA and RFA specimens, respectively.
Further details on specimen preparation and damage characterization are available in the aforementioned study~\cite{malone2021evaluation}.
\begin{figure}[!htb]        
\centering 
\includegraphics[width=1.91in]{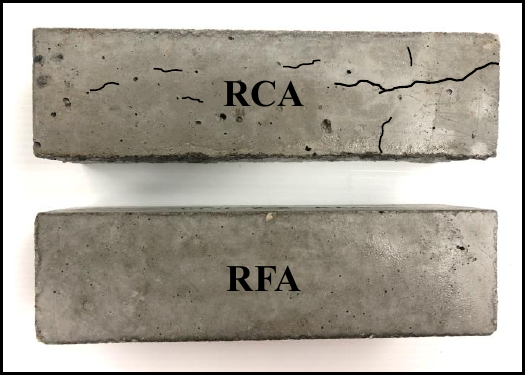}        
\caption{ASR-damaged concrete specimens. Visible cracking is emphasized.}
\label{fig:ASR_specimens} 
\end{figure}
\FloatBarrier
\section{Methods}\label{sec:4}
\subsection{Slow dynamics experimental setup}\label{sec:4.1}
Specimen conditioning was performed using a universal testing machine (Instron 6800 series).
Each specimen was loaded to 6 kN at a rate of 300 N/s and unloaded at 600 N/s.
Prior to the tests reported in this study, all specimens had previously been loaded to 15 kN using a different setup; therefore, the 6 kN conditioning load used here did not exceed the specimens' previous maximum load.
Ultrasonic signals were acquired during three phases (rest, load, and recovery) over a total duration of 40 minutes, as shown in Fig.~\ref{fig:Loading}.

\begin{figure}[!htb]        
\centering 
\includegraphics[width=3in]{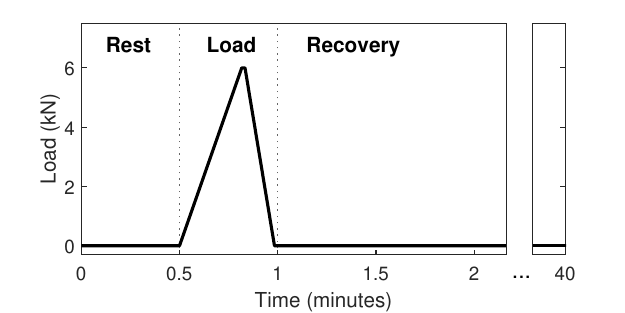}        
\caption{
Loading protocol showing rest, loading, and recovery phases.
}
\label{fig:Loading} 
\end{figure}
The experimental setup for the ultrasonic measurements is shown in Fig.~\ref{fig:TestSetup}.
Two shear transducers (Panametrics V1548, 0.1 MHz) were installed on opposite sides of the specimens and operated in the through-transmission mode.
A high viscosity syrup was used for coupling the transducers.
The transducers were driven by 300 V square pulses (0.1 MHz center frequency) supplied by a pulser/receiver (Olympus 5077PR).
The signals were digitized by an oscilloscope (PicoScope 5444A) with a sampling frequency of 250 MHz and averaged 8 times. 

Thin film RTD sensors (TE Connectivity, 1k$\Omega$, NB-PTCO-126, 4-wire configuration) were surface-mounted on the prism and used to monitor specimen and ambient temperature during testing.
The temperature data was acquired using a DAQ/switch unit (Agilent 34970A) with a multiplexer module (Agilent 34902A).
To minimize the influence of thermal effects during testing, the specimens were wrapped in polyethylene foam and then surrounded by a roll of aluminum insulation, as can be seen in Fig.~\ref{fig:TestSetup}(c).
\begin{figure}[!htb]        
\centering 
\includegraphics[width=4in]{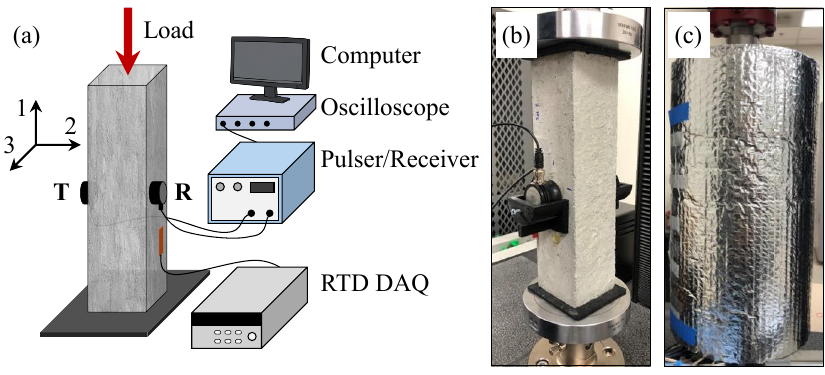}      
\caption{Experimental setup for ultrasonic and temperature measurements.
(a) Schematic of loading, sensors, and data acquisition system.
(b) Photograph of specimen in the test frame without insulation for clarity.
(c) Specimen wrapped in polyethylene foam and aluminum insulation to minimize temperature effects.}
\label{fig:TestSetup} 
\end{figure}
\FloatBarrier
\subsection{Thermal modulation experimental setup}\label{sec:4.2}
To measure the TACs used in the thermal correction process described in Sec.~\ref{sec:2.3}, the concrete specimens were placed in an environmental chamber (Memmert IPP55plus) for a heating and cooling cycle (see Fig.~\ref{fig:TestSetup_thermal}(a)).
During testing, ultrasonic signals were acquired approximately every minute, with the same signal properties as the aforementioned setup.
To monitor the temperature changes during testing, RTDs were placed on the surface of the test specimen, while an additional concrete prism containing an internal thermocouple was placed nearby.
A temperature change rate of 0.5$^\circ$C/hour was implemented to avoid a large thermal gradient.

The temperature history during one thermal modulation test is shown in Fig.~\ref{fig:TestSetup_thermal}(b).
Surface and internal measurements confirmed that thermal gradients were negligible, so the surface RTD measurement was used for TAC analysis, consistent with the measurements available in the slow dynamics setup.
\begin{figure}[!htb]
\centering
\begin{subfigure}{0.35\textwidth}
    \centering
    \includegraphics[width=1.455in]{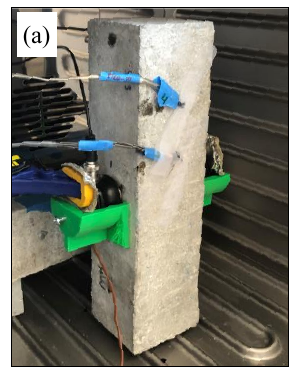} 
\end{subfigure}
\hfill
\begin{subfigure}{0.65\textwidth}
    \centering
    \includegraphics[width=3in]{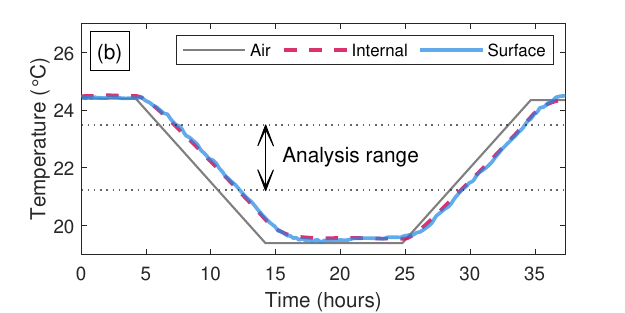}
\end{subfigure}
\caption{Thermal modulation experimental setup.
(a) Specimen in environmental chamber.
(b) Temperature history showing air, internal, and surface measurements with analysis range used for TAC calculation.}
\label{fig:TestSetup_thermal} 
\end{figure}

The analysis range (21.5$^\circ$C to 23.5$^\circ$C) was selected for TAC calculations since it was representative of the room temperature variations experienced during the slow dynamics tests.
Additionally, this selection avoided hysteresis effects near temperature reversals, which did not occur during the monotonic temperature changes of this study (see Sec.~\ref{sec:2.3}).
\FloatBarrier
\subsection{Signal processing}\label{sec:4.3}
A collected ultrasonic signal is shown in Fig.~\ref{fig:Signal}.
The wave type generated during the tests (see Fig.~\ref{fig:TestSetup}) is referred to as $V_{21}$.
This notation describes a shear wave propagating in the ``2" direction (thickness) and polarized in the ``1" direction (parallel to the applied load). 
For shear waves in concrete specimens, $V_{21}$ has been found to be the most sensitive to applied stresses (acoustoelastic effect)~\cite{bompan2018ultrasonic, zhong2022applications}.

For the analysis performed, however, CWI was applied using a time window that included the coda wave.
Because the coda wave field is diffuse and without polarization, CWI does not isolate the velocity change of specific modes, such as $V_{21}$.
Snieder~\cite{snieder2002coda} showed that, after equilibration of the P- and S-wave energies in a Poisson medium, the relative velocity change of the coda wave is a weighted average of the individual relative velocity changes of P- and S-waves (0.09 ${dV/V}_P$ + 0.91 ${dV/V}_S$).
The velocity changes observed in this study are therefore expected to be strongly influenced by the S-wave contribution.
\begin{figure}[!htb]        
\centering 
\includegraphics[width=3in]{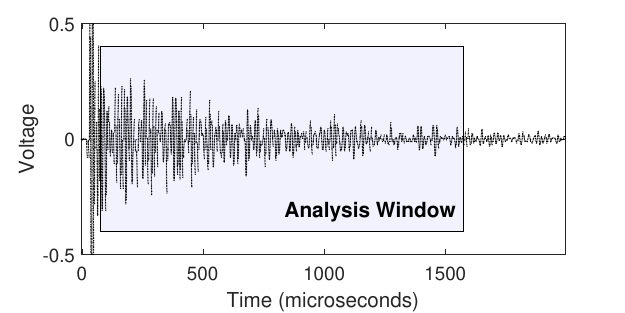}        
\caption{Typical ultrasonic signal with the selected analysis window (shaded) used for CWI processing.}
\label{fig:Signal} 
\end{figure}
\FloatBarrier
\section{Results and discussion}\label{sec:5}
\subsection{Thermal modulation results and temperature correction}\label{sec:5.1}
The results of a thermal modulation test and temperature correction for one prism from Mix 2 ($f'_c$ = 40 MPa) are shown in Fig.~\ref{fig:Temp Correction}.
Fig.~\ref{fig:Temp Correction}(a) compares the linear fits for the heating and cooling phases of the thermal modulation test, while Fig.~\ref{fig:Temp Correction}(b) and (c) present the raw and temperature corrected relative velocity changes from slow dynamics tests.
Despite insulation (see Fig.~\ref{fig:TestSetup}(c)), small temperature fluctuations of approximately 0.2$^\circ$C were observed during testing.
These fluctuations induced a linear drift in the relative velocity change measurements, which resulted in a significant deviation from the expected logarithmic recovery, especially at later times ($> 10^3$ s).
After applying the temperature correction procedure (see Sec.~\ref{sec:2.3}) the drift was removed. Because the observed fluctuations were small ($\sim0.2^\circ$C) and the specimens were insulated, internal-surface thermal gradients were expected to be negligible, consistent with the agreement between surface and internal measurements observed during the thermal modulation tests (Sec.~\ref{sec:4.2}).

Table~\ref{Tab:TMNL_results} summarizes the TACs for all specimen groups.
The cooling and heating TACs were obtained by fitting $dV/V$ versus temperature separately over the cooling and heating phases, respectively.  These phase-specific TACs were then used for temperature correction based on whether the specimen experienced cooling or heating (Eq.~\ref{equation 6}).
The combined TAC was obtained from a linear fit of both cooling and heating data and is reported only as a reference value; it was not used for correction.
The phase-specific approach reflects concrete's hysteretic temperature response (see Sec.~\ref{sec:2.3}).
The correction procedure was applicable for the gradual, limited-range temperature changes encountered in this study, while non-monotonic or rapid temperature changes would likely require a different approach.
Heating TACs exceeded cooling TACs for all groups, consistent with previous thermal modulation studies~\cite{sun2020determination}.
Among the intact specimens, the TACs generally increased with compressive strength.
The ASR-damaged specimens exhibited higher TACs than the intact group, with the RCA specimen having values nearly double those of the RFA specimen.
\begin{figure}[!htb]
\centering
\begin{subfigure}{0.8\textwidth}
\centering
\includegraphics[width=3in]{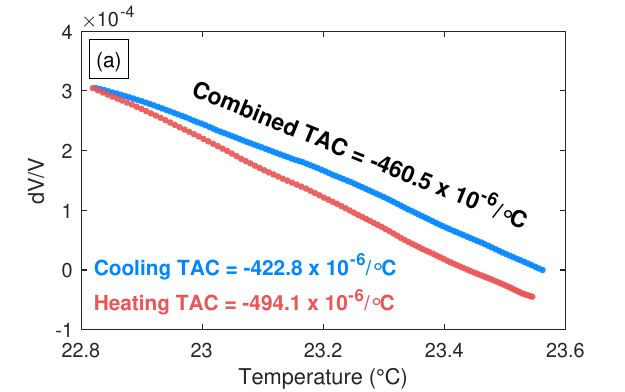}
\end{subfigure}\\%
\bigskip
\begin{subfigure}{0.48\textwidth}
\centering
\includegraphics[width=2.25in]{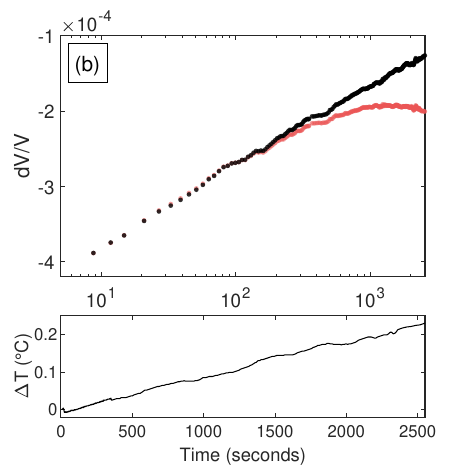}
\end{subfigure}
\begin{subfigure}{0.48\textwidth}
\centering
\includegraphics[width=2.25in]{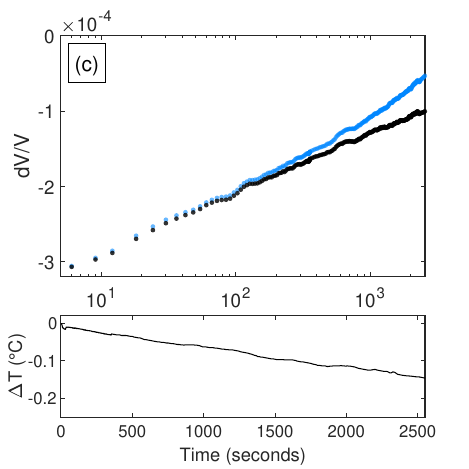}
\end{subfigure}
\caption{Determination of TACs and application of temperature correction (sample from Mix 2, $f'_c$ = 40 MPa).  
(a) Measured $dV/V$ versus temperature during heating (red) and cooling (blue) phases, showing TACs.  
(b) Recovery curve during ambient heating before (red) and after (black) temperature correction.  
(c) Recovery curve during ambient cooling before (blue) and after (black) temperature correction.}
\label{fig:Temp Correction} 
\end{figure}
\begin{table}[htb]
\centering
\caption{TACs for intact and ASR-damaged specimens.} 
{\small
\begin{tabular}{lcccc}
\toprule
Mix             & $f'_c$ & Cooling TAC                  & Heating TAC                 & Combined TAC                \\
                & (MPa)  & ($\times10^{-6}/^{\circ}$C)  & ($\times10^{-6}/^{\circ}$C) & ($\times10^{-6}/^{\circ}$C) \\
\midrule
Mix 1           & 32     & $-441.1$                     & $-520.0$                    & $-496.5$                    \\
Mix 2           & 40     & $-422.8$                     & $-494.1$                    & $-460.5$                    \\
Mix 3           & 49     & $-533.4$                     & $-566.2$                    & $-555.5$                    \\
Mix 4           & 56     & $-596.7$                     & $-679.0$                    & $-643.4$                    \\
\midrule
RFA (ASR)       & 35     & $-681.9$                     & $-951.0$                    & $-773.2$                    \\
RCA (ASR)       & 35     & $-1211.2$                    & $-1517.7$                   & $-1364.5$                   \\
\bottomrule
\end{tabular}}
\label{Tab:TMNL_results}
\end{table} 
\FloatBarrier
\subsection{Compressive loading results}\label{sec:5.2}
\subsubsection{Intact specimens}\label{sec:5.2.1}
Fig.~\ref{fig:Strengths_combined} presents the recovery of the four intact concrete groups ($f'_c =$ 32, 40, 49, and 56 MPa) after an applied compressive force (experimental setup displayed in Fig.~\ref{fig:TestSetup}).
Each group had three specimens, represented by the individual curves, and the group averages are denoted by the bold lines.
Table~\ref{tab:strength_fit_values} lists the mean fit parameters for each group, along with the coefficients of variation (CVs) for the primary mixes.
Parameters for the additional intact specimens (Sec.~\ref{sec:5.2.2}) and the ASR-damaged specimens (Sec.~\ref{sec:5.2.3}) are included for comparison.
The variation between specimens of the same group may reflect natural material differences (in-batch variance) or measurement error.
Recovery parameter CVs exceed the corresponding batch strength CVs (Table~\ref{tab:mix_designs}), reflecting potential measurement variability.
A discussion of experimental repeatability is provided in Appendix~\ref{app:repeatability}.
\begin{figure}[!htb]
\centering
\includegraphics[width=3in]{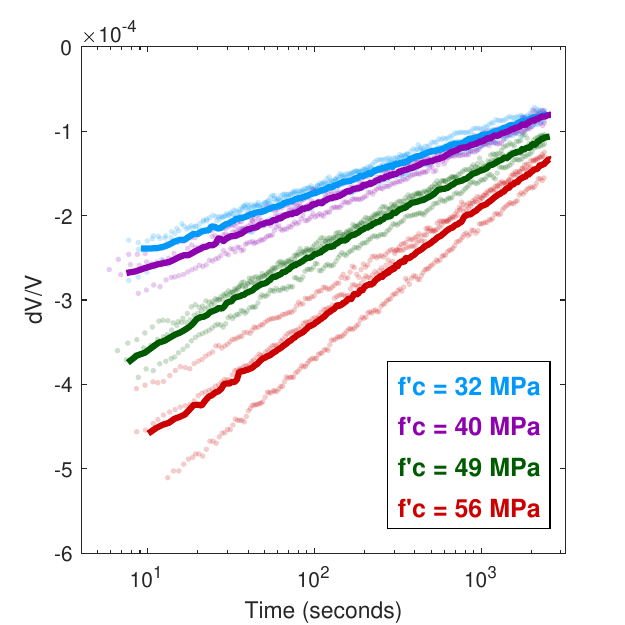}
\caption{Recovery of $dV/V$ following compressive loading for intact concrete mixes with compressive strengths of $f'_c =$ 32, 40, 49, and 56 MPa.
Individual specimen responses are shown as lighter curves, with bold lines indicating group averages.}
\label{fig:Strengths_combined}
\end{figure}

\begin{table}[!htb]
\centering
\caption{Recovery parameters for all specimens, with coefficients of variation for the primary mixes. Compressive strengths ($f'_c$) are 28-day batch values.}
\label{tab:strength_fit_values}
\begin{tabular}{lcS[table-format=2.2]rS[table-format=-2.2]rr}
\toprule
 & $f'_c$ & {$m_v$} & \multirow{2}{*}{CV$_m$} & {$c$} & \multirow{2}{*}{CV$_c$} & $t_r$ \\
 & (MPa) & {($\times10^{-5}/s$)} & & {($\times10^{-4}$)} & & (h) \\
\midrule
\multicolumn{7}{l}{\textit{Primary intact mixes (mean of three specimens)}} \\
Mix 1         & 32 & 6.77  & $6.8$\%  & -3.08  & $7.1$\%  & 9.9 \\
Mix 2         & 40 & 7.56  & $7.9$\%  & -3.38  & $7.5$\%  & 8.2 \\
Mix 3         & 49 & 10.24 & $5.5$\%  & -4.52  & $6.0$\%  & 7.3 \\
Mix 4         & 56 & 13.76 & $14.3$\% & -6.02  & $12.8$\% & 6.6 \\
\midrule
\multicolumn{7}{l}{\textit{Additional intact specimens}} \\
Specimen \#1  & 34 & 9.34  & --       & -3.68  & --       & 2.4 \\
Specimen \#2  & 35 & 11.17 & --       & -4.52  & --       & 3.1 \\
\midrule
\multicolumn{7}{l}{\textit{ASR-damaged specimens}} \\
RFA           & 35 & 19.06 & --       & -11.85 & --       & 458 \\
RCA           & 35 & 28.04 & --       & -15.84 & --       & 124 \\
\bottomrule
\end{tabular}
\end{table}
Both the recovery rate, $m_v$, and the velocity drop magnitude, $|c|$, increased with strength, indicating that the higher strength prisms underwent larger transient softening, but recovered at a faster rate.
Specifically, $|c|$ increased with increasing strength from $3.08 \times 10^{-4}$ for $f'_c = 32$ MPa to $6.02 \times 10^{-4}$ for $f'_c = 56$ MPa, while $m_v$ increased from $6.77 \times 10^{-5}$/s to $13.76 \times 10^{-5}$/s.
As a result, the recovery time $t_r$ shortened from 9.9 h to 6.6 h; thus, the stronger specimens regained stiffness more quickly.

The increase in $|c|$ with compressive strength may appear counterintuitive, as greater nonlinearity in concrete is typically associated with a weaker or damaged microstructure.
For example, Payan et al.~\cite{payan2014quantitative} measured lower hysteretic nonlinearity in high performance concrete than in ordinary concrete, attributing the difference to paste-aggregate adherence and aggregate distribution.
Similarly, Seo et al.~\cite{seo2025utilizing} found that field cores with lower compressive strengths had higher nonlinear parameters, which they interpreted as an index of microstructural damage in the as-built concrete.
Although both studies measured the fast nonlinear response, fast and slow dynamics have been shown to occur together across diverse materials and are believed to share a common physical origin~\cite{johnson2005slow}, so a similar expectation could reasonably extend to the slow dynamics measured here.
In those studies, however, the greater nonlinearity was attributed to differences in composition or accumulated damage that varied alongside strength, whereas the present groups are companion mixes made from the same materials, differing only in mix proportions, with strength varied by design.
Across the groups in this study, strength varies with the water-cement ratio (Table~\ref{tab:mix_designs}), and the lower w/c of the stronger mixes produces a denser cement paste and a stiffer matrix.

The influence of the matrix on slow dynamics can be understood through the hard/soft paradigm of nonlinear mesoscopic elasticity, under which concrete is treated as a rigid matrix containing a small volume of elastically soft constituents~\cite{johnson2005slow}.
Within this framework, the slow dynamics response originates in and is governed by this soft ``bond system" of grain contacts and microcracks, rather than by the rigid matrix itself~\cite{guyer2009nonlinear}.
Because the rigid matrix resists deformation, the applied strain is amplified in the bond system, where local strains can exceed the macroscopic strain by an order of magnitude~\cite{guyer2009nonlinear}.
We propose that the denser, stiffer matrix of the higher strength specimens produces greater strain amplification at the soft bond system under the same conditioning force, resulting in a larger transient softening observed as the larger velocity drop.
Furthermore, the stiffer matrix may also impose larger local restoring stresses on the perturbed contacts after unloading, accelerating their recovery and contributing to the faster recovery rates ($m_v$) of the stronger specimens.
Under this interpretation, the larger velocity drop and faster recovery of the stronger specimens do not indicate a weaker or damaged material, but rather reflect the greater strain concentration imposed on the soft bond system by the surrounding rigid matrix.
Related microstructural factors that evolve with strength, such as pore structure, moisture state, and crack volume, may also contribute.
As proposed by Bittner and Popovics~\cite{bittner2022transient}, such variations in crack volume and internal moisture, rather than the base microstructure, are key factors governing the slow dynamic behavior.
Further testing is needed to isolate the relative roles of these factors.
\FloatBarrier
\subsubsection{Additional intact specimens}\label{sec:5.2.2}
To investigate the generality of the trends observed in Sec.~\ref{sec:5.2.1}, two additional specimens (details provided in Sec.~\ref{sec:3.2}) were tested under the same conditions.
Their recovery curves are plotted in Fig.~\ref{fig:comparison_with_others} against 95\% prediction intervals (colored bands) formed from the fit parameters of the four primary mixes (see Table~\ref{tab:strength_fit_values}).
\begin{figure}[!htb]        
\centering 
\includegraphics[width=3in]{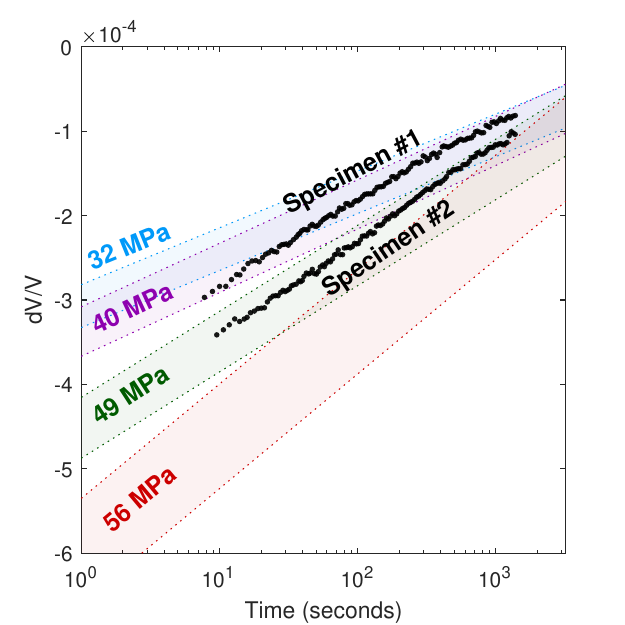}        
\caption{Recovery curves for two additional intact specimens ($f'_c =$ 34 and 35 MPa) overlaid on 95\% prediction intervals (shaded bands) from the four primary strength groups.}
\label{fig:comparison_with_others} 
\end{figure}

The fit parameters for both specimens are listed in Table~\ref{tab:strength_fit_values}.
Specimen \#1 ($f'_c$ = 34 MPa) had a recovery rate of $m_v=9.34\times10^{-5}/s$ and a velocity drop of $|c|=3.68\times10^{-4}$ ($t_r=2.4$ h).
Specimen \#2 ($f'_c$ = 35 MPa) had a recovery rate of $m_v=11.17\times10^{-5}/s$ and a velocity drop of $|c|=4.52\times10^{-4}$ ($t_r=3.1$ h).
The two specimens shared similar recovery times, shorter than all four primary groups, but their recovery behavior aligned more closely with the $f'_c = 40\sim49$ MPa groups than with the $f'_c = 32$ MPa group.
The behavior indicates that recovery parameters are sensitive to age-dependent changes in the microstructure.

This behavior may be explained by continued hydration during the five-year storage period, which likely increased the specimens' present strength beyond their 28-day values and aligned their recovery behavior more closely with the higher-strength groups.
Moisture may also play a role, as lower saturation has been shown to increase hysteretic nonlinearity in concrete~\cite{payan2010effect}.
The prolonged storage of these older specimens could have reduced their moisture content.
Specimens of similar age stored in the same conditions had a measured moisture content of 2.5\% by mass.
Beyond these age-dependent changes to the pore structure and internal moisture, different aggregate properties and mix designs may also contribute, since grain size has been shown to control characteristic relaxation times~\cite{kober2022material}.
\FloatBarrier
\subsubsection{ASR-damaged specimens}\label{sec:5.2.3}
Two ASR-damaged specimens were tested and their responses compared with the intact specimens.
Details of the ASR-damaged specimens and their preparation are provided in Sec.~\ref{sec:3.3}.
The slow dynamics responses are shown alongside the four intact strength groups in Fig.~\ref{fig:Damaged}.

Within the intact specimen groups, we found that stronger mixes had larger velocity drops when undergoing the same conditioning.
However, damaged specimens are generally known to display similar behavior, with increased nonlinearity and sensitivity to conditioning.
For example, Larose et al.~\cite{larose2013ultrasonic} observed that the relative velocity drop was considerably larger in mechanically damaged concrete than in undamaged companion samples.
Similarly, ASR damage has been shown to alter the slow dynamics response, with the viscosity of gel within the cracks introducing time-dependent creep behavior during conditioning~\cite{kodjo2011impact}, and increased responses have been reported in thermally damaged asphalt concrete specimens, where the change in the recovery rate was used as a damage indicator~\cite{bekele2017slow}.

\begin{figure}[!htb]        
\centering 
\includegraphics[width=3in]{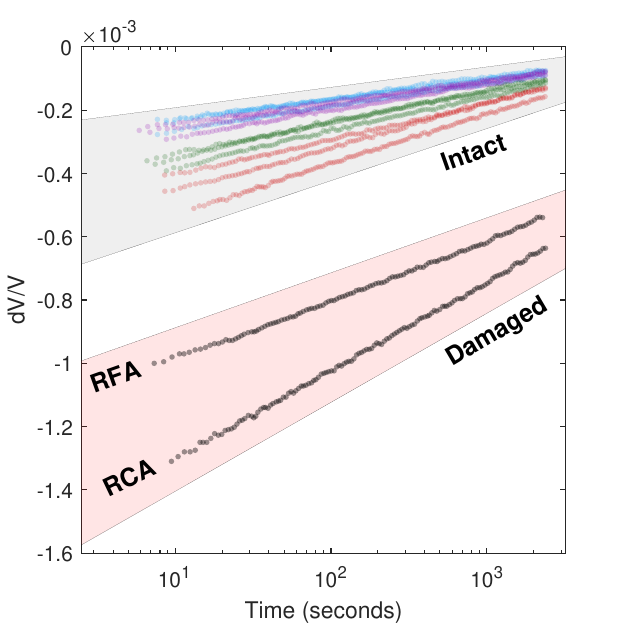}        
\caption{Recovery curves for ASR-damaged specimens (RCA and RFA) compared with the four intact strength groups.}
\label{fig:Damaged} 
\end{figure}

Consistent with those studies, our damaged specimens had different recovery behavior than the intact specimens.
ASR-damaged specimens had velocity drops of $|c|=15.84\times10^{-4}$ (RCA) and $|c|=11.85\times10^{-4}$ (RFA), exceeding all intact groups ($|c|=3.08\times10^{-4}$ to $6.02\times10^{-4}$).
Their recovery rates ($m_v=28.04\times10^{-5}/s$ for RCA and $m_v=19.06\times10^{-5}/s$ for RFA) were also faster than the intact groups ($m_v=6.77\times10^{-5}/s$ to $m_v=13.76\times10^{-5}/s$) (Table~\ref{tab:strength_fit_values}).
Despite the faster recovery rates, their recovery times were considerably longer, with $t_r=124$ h for the RCA specimen and $t_r=458$ h for the RFA specimen.
These long recovery times reflect a departure from the trend observed in the intact specimens, where larger velocity drops were accompanied by proportionally faster recoveries, resulting in relatively consistent recovery times ($t_r=9.9$ h for $f'_c$ = 32 MPa vs. $t_r=6.6$ h for $f'_c$ = 56 MPa).
For thermal damage, Larose et al.~\cite{larose2013ultrasonic} likewise found that increasing damage produced both larger velocity drops and faster recovery rates.

In the ASR-damaged specimens, by contrast, the softening increased without a proportional increase in recovery rate.
Longer recovery times have also been observed in mechanically damaged concrete, where Bentahar et al.~\cite{bentahar2006hysteretic} found a 60\% increase in damaged samples relative to undamaged samples, less than the increase measured for the ASR-damaged specimens in this study.
Kodjo et al.~\cite{kodjo2011impact} showed that gel-filled cracks lengthened the creep response time during conditioning by a factor of more than three relative to empty cracks, attributing the delayed response to the viscosity of the gel.
The loss of proportionality observed in the present study may therefore reflect a mechanism associated with specific damage types, such as ASR, rather than a universal response of damaged concrete.

The RFA specimen, despite having less length change (0.0230\%) than the RCA specimen (0.0721\%), had a recovery time over three times longer.
In our previous study~\cite{malone2021evaluation}, scanning electron microscopy (SEM) revealed that damage in the RFA specimen was characterized by many isolated microcracks confined to individual sand grains, only partially extending into the paste and not yet forming an interconnected network.
These microcracks were mostly empty, with only small amounts of gel observed near the paste-aggregate interface.
The RCA specimen had more advanced damage, with microcracks bridging between aggregates to form an interconnected network, and gel found within cracks and adjacent voids.
When measured at the same expansion level, resonant frequency testing showed no frequency drop in the RFA specimen, while the RCA specimen had a reduction of more than 10\%.
Thus, it was concluded that the RCA specimen's larger, interconnected cracks resulted in a measurable decrease in its elastic modulus, whereas the RFA specimen's isolated microcracks had less impact on the overall matrix and thus the modulus.

These observations provide context for the slow dynamics results observed in Fig.~\ref{fig:Damaged}.
Although gel was more prevalent in the RCA specimen, the RFA specimen had the longer recovery time, indicating that the behavior of the two specimens may not be governed by gel alone.
We believe that the distributed microcracking in the RFA specimen contributed to its longer relaxation period.
While the widespread microcracking may have resulted in less linear expansion, the larger affected volume resulted in a slower return to equilibrium.
Further investigation is needed to clarify the influence of alkali-silica gel and the presence of gel-filled cracks on slow dynamics.
\FloatBarrier
\section{Conclusions}\label{sec:6}
In this study, we investigated the slow dynamic behavior of concrete specimens with varying strengths and ASR damage.
The specimens were conditioned by a compressive load and the subsequent wave velocity recovery was monitored.
A self-referencing temperature correction procedure was used to correct the influence of temperature fluctuations on velocity measurements.

Major findings from the experimental studies are listed below:
\begin{enumerate}
    \item
    Thermo-acoustic coefficients (TACs) were determined by thermal modulation tests and used to correct the effect of temperature during slow dynamics experiments.
    The proposed temperature correction procedure removed the relative velocity change drift caused by ambient cooling or heating (typically $\pm0.2^{\circ}$C).
    For intact specimens, higher TACs were primarily observed in the higher-strength groups, though the increase was not uniform across all mix designs.
    Furthermore, ASR-damaged specimens exhibited significantly larger TACs than the intact group.
    
    \item
    For intact concrete, the recovery rate, $m_v$, and velocity drop magnitude, $|c|$, increased with compressive strength.
    That is, despite larger softening, higher strength specimens returned faster to their preconditioned stiffness ($t_r = 9.9$ h for $f'_c = 32$ MPa vs. $t_r = 6.6$ h for $f'_c = 56$ MPa).
    
    \item
    To evaluate the generality of these trends, two additional, older intact specimens were tested.
    While their overall recovery behavior remained characteristic of intact concrete, their specific recovery parameters fell outside the expected strength prediction intervals of the primary groups.
    This deviation is likely due to age and moisture dependent microstructure changes.
    This indicates the need for future work that considers these effects when comparing recovery parameters across sets of specimens.
    
    \item
    ASR-damaged specimens had increased softening without proportionally faster recovery rates, resulting in longer recovery times ($t_r$ up to 458 h).
    Notably, recovery took over three times longer for the RFA specimen than the RCA specimen, despite its lower expansion level.
    This indicates that recovery behavior is influenced not only by the severity of damage, but also by the underlying damage mechanisms.
    In this case, the longer recovery time of the RFA specimen is attributed to its widespread, isolated microcracking acting over a larger volume of the matrix, in contrast to the interconnected macrocrack network concentrated at the coarse aggregates in the RCA specimen.
    
    \item
    The temperature-corrected slow dynamics measurements in this study were performed under controlled laboratory conditions on concrete prism specimens, with conditioning applied using a testing machine. Therefore, future work should evaluate the robustness of the methodology under field-representative conditions, environmental fluctuations, and practical methods of loading application.
\end{enumerate}
\begin{appendices}
\renewcommand{\thefigure}{\arabic{figure}}
\renewcommand{\thetable}{\arabic{table}}
\setcounter{figure}{12}
\setcounter{table}{4}
\section{Experimental repeatability}\label{app:repeatability}
To evaluate the repeatability of the measurements, four consecutive recovery tests were conducted on the same intact specimen (Mix 1, $f'_c$ = 32 MPa) using the primary compressive loading setup (Fig.~\ref{fig:TestSetup}).
A five-hour recovery period was kept between tests to allow the specimen to return near its initial state before retesting.
The results of the four runs are summarized in Fig.~\ref{fig:Repeat} and Table~\ref{tab:test_results}.

Across the four tests, the recovery rate ($m_v$) had a CV of 4.3\%, while the velocity drop magnitude ($|c|$) had a CV of 3.6\%.
Prior to applying temperature correction (Section~\ref{sec:2.3}), the CVs were 16.1\% and 5.0\%, respectively.
This demonstrates that reliable assessment of recovery behavior requires careful temperature control and correction, particularly when evaluating the recovery rate.
\begin{figure}[!htb]
\centering
\includegraphics[width=2.5in]{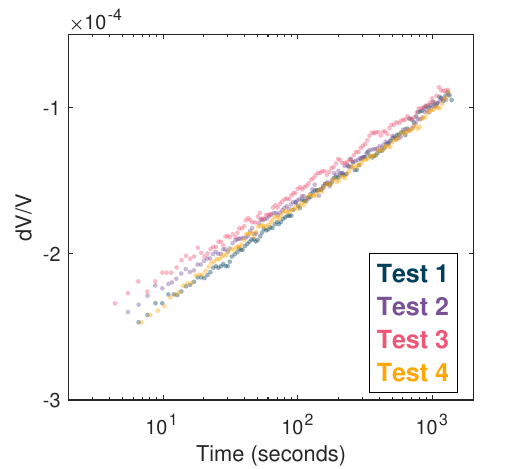}
\caption{Recovery of $dV/V$ over four repeated tests on the same intact specimen (Mix 1, $f'_c$ = 32 MPa).}
\label{fig:Repeat}
\end{figure}
\begin{table}[!htb]
\centering
\caption{Repeatability of recovery parameters from four consecutive tests (Mix 1, $f'_c$ = 32 MPa).}
\label{tab:test_results}
\begin{tabular}{lccc}
\toprule
Test & $m_v$               & $c$                   & $t_r$   \\
(\#) & ($\times10^{-5}/s$) & ($\times10^{-4}$)     & (hours) \\
\midrule
1    & $6.79$              & $-3.04$               & $8.3$     \\
2    & $6.27$              & $-2.89$               & $11.2$    \\
3    & $6.16$              & $-2.79$               & $9.4$     \\
4    & $6.45$              & $-2.96$               & $10.9$    \\
\midrule
Mean & $6.42$              & $-2.92$               & $9.9$     \\
CV   & $4.3$\%             & $3.6$\%               & $13.5$\%  \\
\bottomrule
\end{tabular}
\end{table}
\end{appendices}

\section*{Acknowledgments}
This research is supported by the U.S. Department of Energy – Nuclear Energy University Program (NEUP) under the contract DE-NE0008544. 

\section*{Declarations}
\begin{itemize}
\item Funding -
U.S. Department of Energy DE-NE0008544. 
\item Conflict of interest -
The corresponding author Jinying Zhu is an associate editor of Journal of Nondestructive Evaluation. 
\item Ethics approval and consent to participate  -
Not applicable
\item Consent for publication -
The publisher has the authors' permission to publish this paper and the research findings within it.
\item Data availability -
The data that support the findings of this study are available from the corresponding author J.Z. upon reasonable request.
\item Materials availability -
Materials are available upon reasonable request.
\item Code availability  -
The code is available upon reasonable request.
\item Author contributions -
J.Z. and C.M. conceptualized the methodology. C.M. collected data and performed analysis. All authors wrote, edited, and reviewed the manuscript.
\end{itemize}

\bibliography{references}

@TechReport{ASTM-C39-24,
   Author       = {{ASTM International}},
   Title        = {Standard Test Method for Compressive Strength of Cylindrical Concrete Specimens},
   Type         = "{ASTM C39-24}",
   Institution  = {ASTM International},
   Address      = {West Conshohocken, PA},
   Year         = {2024}
}

@article{ahmed2003effect,
  title = {The Effect of Alkali Reactivity on the Mechanical Properties of Concrete},
  author = {Ahmed, Tarig and Burley, Eldon and Rigden, Stephen and {Abu-Tair}, Abid I.},
  year = {2003},
  journal = {Construction and Building Materials},
  volume = {17},
  number = {2},
  issn = {09500618},
  doi = {10.1016/S0950-0618(02)00009-0},
  urldate = {2025-10-01},
  copyright = {https://www.elsevier.com/tdm/userlicense/1.0/},
  langid = {english}
}

@article{bekele2017slow,
    author = {Bekele, Abiy and Birgisson, Björn and Ryden, Nils and Gudmarsson, Anders},
    title = {Slow dynamic diagnosis of asphalt concrete specimen to determine level of damage caused by static low temperature conditioning},
    journal = {AIP Conference Proceedings},
    volume = {1806},
    number = {1},
    pages = {080012},
    year = {2017},
    issn = {0094-243X},
    doi = {10.1063/1.4974637}
}

@article{bentahar2006hysteretic,
  title = {Hysteretic Elasticity in Damaged Concrete: {{Quantitative}} Analysis of Slow and Fast Dynamics},
  shorttitle = {Hysteretic Elasticity in Damaged Concrete},
  author = {Bentahar, M. and El Aqra, H. and El Guerjouma, R. and Griffa, M. and Scalerandi, M.},
  year = {2006},
  journal = {Physical Review B},
  volume = {73},
  number = {1},
  issn = {1098-0121, 1550-235X},
  doi = {10.1103/PhysRevB.73.014116},
  urldate = {2024-04-17},
  copyright = {http://link.aps.org/licenses/aps-default-license},
  langid = {english}
}

@article{bissig2003intermittent,
  title={Intermittent dynamics and hyper-aging in dense colloidal gels},
  author={Bissig, Hugo and Romer, Sara and Cipelletti, Luca and Trappe, Veronique and Schurtenberger, Peter},
  journal={PhysChemComm},
  volume={6},
  number={5},
  year={2003},
  issn = {14602733},
  doi = {10.1039/b211806h},
  urldate = {2025-10-01},
  langid = {english}
}

@article{bittner2019direct,
  title = {Direct Imaging of Moisture Effects during Slow Dynamic Nonlinearity},
  author = {Bittner, J. A. and Popovics, J. S.},
  year = {2019},
  journal = {Applied Physics Letters},
  volume = {114},
  number = {2},
  issn = {0003-6951, 1077-3118},
  doi = {10.1063/1.5063904},
  urldate = {2024-04-17},
  langid = {english}
}

@article{bittner2021mechanistic,
  title = {Mechanistic Diffusion Model for Slow Dynamic Behavior in Materials},
  author = {Bittner, J.A. and Popovics, J.S.},
  year = {2021},
  journal = {Journal of the Mechanics and Physics of Solids},
  volume = {150},
  issn = {00225096},
  doi = {10.1016/j.jmps.2021.104355},
  urldate = {2025-10-01},
  langid = {english}
}

@article{bittner2022transient,
  title = {Transient Nonlinear Vibration Characterization of Building Materials in Sequential Impact Scale Experiments},
  author = {Bittner, James A. and Popovics, John S.},
  year = {2022},
  journal = {Frontiers in Built Environment},
  volume = {8},
  issn = {2297-3362},
  doi = {10.3389/fbuil.2022.949484},
  urldate = {2024-03-07},
  langid = {english}
}

@article{bompan2018ultrasonic,
  title = {Ultrasonic Tests in the Evaluation of the Stress Level in Concrete Prisms Based on the Acoustoelasticity},
  author = {Bompan, Karen F. and Haach, Vladimir G.},
  year = {2018},
  journal = {Construction and Building Materials},
  volume = {162},
  issn = {09500618},
  doi = {10.1016/j.conbuildmat.2017.11.153},
  urldate = {2024-02-20},
  langid = {english}
}

@article{chen2010rapid,
  title = {Rapid Evaluation of Alkali--Silica Reactivity of Aggregates Using a Nonlinear Resonance Spectroscopy Technique},
  author = {Chen, Jun and Jayapalan, Amal R. and Kim, Jin-Yeon and Kurtis, Kimberly E. and Jacobs, Laurence J.},
  year = {2010},
  journal = {Cement and Concrete Research},
  volume = {40},
  number = {6},
  issn = {00088846},
  doi = {10.1016/j.cemconres.2010.01.003},
  urldate = {2024-01-24},
  langid = {english}
}

@article{das2012implication,
  title = {Implication of Pore Size Distribution Parameters on Compressive Strength, Permeability and Hydraulic Diffusivity of Concrete},
  author = {Das, B.B. and Kondraivendhan, B.},
  year = {2012},
  journal = {Construction and Building Materials},
  volume = {28},
  number = {1},
  issn = {09500618},
  doi = {10.1016/j.conbuildmat.2011.08.055},
  urldate = {2025-10-01},
  copyright = {https://www.elsevier.com/tdm/userlicense/1.0/},
  langid = {english}
}

@article{guyer1995hysteresis,
  title = {Hysteresis, {{Discrete Memory}}, and {{Nonlinear Wave Propagation}} in {{Rock}}: {{A New Paradigm}}},
  shorttitle = {Hysteresis, {{Discrete Memory}}, and {{Nonlinear Wave Propagation}} in {{Rock}}},
  author = {Guyer, R. A. and McCall, K. R. and Boitnott, G. N.},
  year = {1995},
  journal = {Physical Review Letters},
  volume = {74},
  number = {17},
  issn = {0031-9007, 1079-7114},
  doi = {10.1103/PhysRevLett.74.3491},
  urldate = {2025-09-21},
  copyright = {http://link.aps.org/licenses/aps-default-license},
  langid = {english}
}

@article{guyer1999nonlinear,
  title = {Nonlinear {{Mesoscopic Elasticity}}: {{Evidence}} for a {{New Class}} of {{Materials}}},
  shorttitle = {Nonlinear {{Mesoscopic Elasticity}}},
  author = {Guyer, Robert A. and Johnson, Paul A.},
  year = {1999},
  journal = {Physics Today},
  volume = {52},
  number = {4},
  issn = {0031-9228, 1945-0699},
  doi = {10.1063/1.882648},
  urldate = {2025-09-21},
  langid = {english}
}

@book{guyer2009nonlinear,
  title = {Nonlinear {{Mesoscopic Elasticity}}: {{The Complex Behaviour}} of {{Granular Media}} Including {{Rocks}} and {{Soil}}},
  shorttitle = {Nonlinear {{Mesoscopic Elasticity}}},
  author = {Guyer, Robert A. and Johnson, Paul A.},
  year = {2009},
  edition = {1},
  publisher = {Wiley},
  doi = {10.1002/9783527628261},
  urldate = {2024-01-24},
  isbn = {978-3-527-40703-3 978-3-527-62826-1},
  langid = {english}
}

@article{johnson1996resonance,
  title = {Resonance and Elastic Nonlinear Phenomena in Rock},
  author = {Johnson, Paul A. and Zinszner, Bernard and Rasolofosaon, Patrick N. J.},
  year = {1996},
  journal = {Journal of Geophysical Research: Solid Earth},
  volume = {101},
  number = {B5},
  issn = {0148-0227},
  doi = {10.1029/96JB00647},
  urldate = {2024-02-19},
  langid = {english}
}

@article{johnson2005slow,
  title = {Slow Dynamics and Anomalous Nonlinear Fast Dynamics in Diverse Solids},
  author = {Johnson, Paul and Sutin, Alexander},
  year = {2005},
  journal = {The Journal of the Acoustical Society of America},
  volume = {117},
  number = {1},
  issn = {0001-4966, 1520-8524},
  doi = {10.1121/1.1823351},
  urldate = {2024-01-23},
  langid = {english}
}

@article{kim2018situ,
  title = {In Situ Nonlinear Ultrasonic Technique for Monitoring Microcracking in Concrete Subjected to Creep and Cyclic Loading},
  author = {Kim, Gun and Loreto, Giovanni and Kim, Jin-Yeon and Kurtis, Kimberly E. and Wall, James J. and Jacobs, Laurence J.},
  year = {2018},
  journal = {Ultrasonics},
  volume = {88},
  issn = {0041624X},
  doi = {10.1016/j.ultras.2018.03.006},
  urldate = {2025-10-01},
  langid = {english}
}

@article{kober2022material,
  title = {Material {{Grain Size Determines Relaxation-Time Distributions}} in {{Slow-Dynamics Experiments}}},
  author = {Kober, J. and Gliozzi, A.S. and Scalerandi, M. and Tortello, M.},
  year = {2022},
  journal = {Physical Review Applied},
  volume = {17},
  number = {1},
  publisher = {American Physical Society (APS)},
issn = {2331-7019},
  doi = {10.1103/physrevapplied.17.014002},
  urldate = {2025-07-24},
  copyright = {https://link.aps.org/licenses/aps-default-license},
  langid = {english}
}

@article{kodjo2011impact,
  title = {Impact of the Alkali--Silica Reaction Products on Slow Dynamics Behavior of Concrete},
  author = {Kodjo, Apedovi S. and Rivard, Patrice and {Cohen-Tenoudji}, Frederic and Gallias, Jean-Louis},
  year = {2011},
  journal = {Cement and Concrete Research},
  volume = {41},
  number = {4},
  issn = {00088846},
  doi = {10.1016/j.cemconres.2011.01.011},
  urldate = {2024-02-19},
  langid = {english}
}

@inproceedings{larose2013ultrasonic,
  title = {Ultrasonic Slow Dynamics to Probe Concrete Aging and Damage},
  booktitle = {{{Review of Progress in Quantitative Nondestructive Evaluation}}: {{Volume}} 32},
  author = {Larose, E. and Tremblay, N. and Payan, C. and Garnier, V. and Rossetto, V.},
  year = {2013},
  pages = {1317-1324},
  address = {Denver, Colorado, USA},
  doi = {10.1063/1.4789195},
  urldate = {2024-01-23},
  langid = {english}
}

@article{lee2024slow,
  title = {Slow Dynamic Elasticity at Short Times},
  author = {Lee, SangMin and Weaver, Richard L.},
  year = {2024},
  journal = {Physical Review E},
  volume = {109},
  number = {6},
  issn = {2470-0045, 2470-0053},
  doi = {10.1103/PhysRevE.109.065002},
  urldate = {2025-08-04},
  langid = {english}
}

@article{lesnicki2011characterization,
  title = {Characterization of {{ASR}} Damage in Concrete Using Nonlinear Impact Resonance Acoustic Spectroscopy Technique},
  author = {Le{\'s}nicki, Krzysztof J. and Kim, Jin-Yeon and Kurtis, Kimberly E. and Jacobs, Laurence J.},
  year = {2011},
  journal = {NDT \& E International},
  volume = {44},
  number = {8},
  issn = {09638695},
  doi = {10.1016/j.ndteint.2011.07.010},
  urldate = {2024-01-24},
  langid = {english}
}

@article{lobkis2003codawave,
  title = {Coda-{{Wave Interferometry}} in {{Finite Solids}}: {{Recovery}} of {{P}} -to- {{S Conversion Rates}} in an {{Elastodynamic Billiard}}},
  shorttitle = {Coda-{{Wave Interferometry}} in {{Finite Solids}}},
  author = {Lobkis, Oleg I. and Weaver, Richard L.},
  year = {2003},
  journal = {Physical Review Letters},
  volume = {90},
  number = {25},
  issn = {0031-9007, 1079-7114},
  doi = {10.1103/PhysRevLett.90.254302},
  urldate = {2025-10-01},
  copyright = {http://link.aps.org/licenses/aps-default-license},
  langid = {english}
}

@article{lobkis2009larsen,
  title = {On the {{Larsen}} Effect to Monitor Small Fast Changes in Materials},
  author = {Lobkis, Oleg I. and Weaver, Richard L.},
  year = {2009},
  journal = {The Journal of the Acoustical Society of America},
  volume = {125},
  number = {4},
  issn = {0001-4966, 1520-8524},
  doi = {10.1121/1.3081530},
  urldate = {2024-03-07},
  langid = {english}
}

@article{malone2021evaluation,
  title = {Evaluation of Alkali--Silica Reaction Damage in Concrete Using Linear and Nonlinear Resonance Techniques},
  author = {Malone, Clayton and Zhu, Jinying and Hu, Jiong and Snyder, April and Giannini, Eric},
  year = {2021},
  journal = {Construction and Building Materials},
  volume = {303},
  issn = {09500618},
  doi = {10.1016/j.conbuildmat.2021.124538},
  urldate = {2024-04-15},
  langid = {english}
}

@article{matthews2026advancing,
  title = {Advancing Non-Destructive Concrete Compressive Strength Estimation: {{Large-Scale}} Datasets and Machine Learning Framework},
  shorttitle = {Advancing Non-Destructive Concrete Compressive Strength Estimation},
  author = {Matthews, Benjamin and Allaix, Diego and Wijte, Simon and Vullings, Marcel},
  year = {2026},
  journal = {NDT \& E International},
  volume = {158},
issn = {09638695},
  doi = {10.1016/j.ndteint.2025.103549},
  urldate = {2025-10-01},
  langid = {english}
}

@inproceedings{niederleithinger2013influence,
  title = {Influence of Small Temperature Variations on the Ultrasonic Velocity in Concrete},
  booktitle = {{{Review of Progress in Quantitative Nondestructive Evaluation}}: {{Volume}} 32},
  author = {Niederleithinger, E. and Wunderlich, C.},
  year = {2013},
  pages = {390-397},
    address = {Denver, Colorado, USA},
  doi = {10.1063/1.4789074},
  urldate = {2025-10-01},
  langid = {english}
}

@article{niederleithinger2018processing,
  title = {Processing {{Ultrasonic Data}} by {{Coda Wave Interferometry}} to {{Monitor Load Tests}} of {{Concrete Beams}}},
  author = {Niederleithinger, Ernst and Wang, Xin and Herbrand, Martin and M{\"u}ller, Matthias},
  year = {2018},
  journal = {Sensors},
  volume = {18},
  number = {6},
  issn = {1424-8220},
  doi = {10.3390/s18061971},
  urldate = {2025-10-01},
  langid = {english}
}

@article{payan2007applying,
  title = {Applying Nonlinear Resonant Ultrasound Spectroscopy to Improving Thermal Damage Assessment in Concrete},
  author = {Payan, C. and Garnier, V. and Moysan, J. and Johnson, P. A.},
  year = {2007},
  journal = {The Journal of the Acoustical Society of America},
  volume = {121},
  number = {4},
  issn = {0001-4966},
  doi = {10.1121/1.2710745},
  urldate = {2025-10-01},
  langid = {english}
}

@article{payan2010effect,
  title = {Effect of Water Saturation and Porosity on the Nonlinear Elastic Response of Concrete},
  author = {Payan, C{\'e}dric and Garnier, Vincent and Moysan, Joseph},
  year = {2010},
  journal = {Cement and Concrete Research},
  volume = {40},
  number = {3},
  issn = {00088846},
  doi = {10.1016/j.cemconres.2009.10.021},
  urldate = {2025-10-01},
  langid = {english}
}

@article{pucinotti2013assessment,
  title = {Assessment of in Situ Characteristic Concrete Strength},
  author = {Pucinotti, R.},
  year = {2013},
  journal = {Construction and Building Materials},
  volume = {44},
  issn = {09500618},
  doi = {10.1016/j.conbuildmat.2013.02.041},
  urldate = {2025-10-01},
  langid = {english}
}

@article{riviere2013pump,
  title = {Pump and Probe Waves in Dynamic Acousto-Elasticity: {{Comprehensive}} Description and Comparison with Nonlinear Elastic Theories},
  shorttitle = {Pump and Probe Waves in Dynamic Acousto-Elasticity},
  author = {Rivi{\`e}re, J. and Renaud, G. and Guyer, R. A. and Johnson, P. A.},
  year = {2013},
  journal = {Journal of Applied Physics},
  volume = {114},
  number = {5},
  issn = {0021-8979, 1089-7550},
  doi = {10.1063/1.4816395},
  urldate = {2024-03-07},
  langid = {english}
}

@article{shokouhi2017dynamic,
  title = {Dynamic Acousto-Elastic Testing of Concrete with a Coda-Wave Probe: Comparison with Standard Linear and Nonlinear Ultrasonic Techniques},
  shorttitle = {Dynamic Acousto-Elastic Testing of Concrete with a Coda-Wave Probe},
  author = {Shokouhi, Parisa and Rivi{\`e}re, Jacques and Lake, Colton R. and Le Bas, Pierre-Yves and Ulrich, T.J.},
  year = {2017},
  journal = {Ultrasonics},
  volume = {81},
  issn = {0041624X},
  doi = {10.1016/j.ultras.2017.05.010},
  urldate = {2024-03-07},
  langid = {english}
}

@article{shokouhi2017slow,
  title = {Slow Dynamics of Consolidated Granular Systems: {{Multi-scale}} Relaxation},
  shorttitle = {Slow Dynamics of Consolidated Granular Systems},
  author = {Shokouhi, Parisa and Rivi{\`e}re, Jacques and Guyer, Robert A. and Johnson, Paul A.},
  year = {2017},
  journal = {Applied Physics Letters},
  volume = {111},
  number = {25},
  issn = {0003-6951, 1077-3118},
  doi = {10.1063/1.5010043},
  urldate = {2024-01-22},
  langid = {english}
}

@article{snieder2002coda,
  title = {Coda Wave Interferometry and the Equilibration of Energy in Elastic Media},
  author = {Snieder, Roel},
  year = {2002},
  journal = {Physical Review E},
  volume = {66},
  number = {4},
  pages = {046615},
  issn = {1063-651X, 1095-3787},
  doi = {10.1103/PhysRevE.66.046615},
  urldate = {2025-10-03},
  copyright = {http://link.aps.org/licenses/aps-default-license},
  langid = {english}
}

@article{snieder2017time,
  title = {The Time Dependence of Rock Healing as a Universal Relaxation Process, a Tutorial},
  author = {Snieder, Roel and {Sens-Sch{\"o}nfelder}, Christoph and Wu, Renjie},
  year = {2017},
  journal = {Geophysical Journal International},
  volume = {208},
  number = {1},
  issn = {0956-540X, 1365-246X},
  doi = {10.1093/gji/ggw377},
  urldate = {2024-03-07},
  langid = {english}
}

@article{sun2019thermal,
  title = {Thermal Modulation of Nonlinear Ultrasonic Wave for Concrete Damage Evaluation},
  author = {Sun, Hongbin and Zhu, Jinying},
  year = {2019},
  journal = {The Journal of the Acoustical Society of America},
  volume = {145},
  number = {5},
pages = {EL405–EL409},
  issn = {0001-4966, 1520-8524},
  doi = {10.1121/1.5108532},
  urldate = {2025-10-01},
  langid = {english}
}

@article{sun2020determination,
  title = {Determination of Acoustic Nonlinearity Parameters Using Thermal Modulation of Ultrasonic Waves},
  author = {Sun, Hongbin and Zhu, Jinying},
  year = {2020},
  journal = {Applied Physics Letters},
  volume = {116},
  number = {24},
  issn = {0003-6951, 1077-3118},
  doi = {10.1063/5.0014975},
  urldate = {2024-05-01},
  langid = {english}
}

@phdthesis{sun2020thesis,
  author = {Sun, Hongbin},
  title  = {Thermal Modulation of Nonlinear Ultrasonic Waves},
  school = {University of Nebraska--Lincoln},
  year   = {2020},
  address = {Lincoln, NE},
  type   = {Ph.D. thesis},
}

@article{taffese2025machine,
  title = {Machine Learning in Concrete Durability: Challenges and Pathways Identified by {{RILEM TC}} 315-{{DCS}} towards Enhanced Predictive Models},
  shorttitle = {Machine Learning in Concrete Durability},
  author = {Taffese, Woubishet Zewdu and Hilloulin, Beno{\^i}t and Villagran Zaccardi, Yury and Marani, Afshin and Nehdi, Moncef L. and Hanif, Muhammad Usman and Kamath, Muralidhar and Nunes, Sandra and {Von Greve-Dierfeld}, Stefanie and Kanellopoulos, Antonios},
  year = {2025},
  journal = {Materials and Structures},
  volume = {58},
  number = {4},
  issn = {1359-5997, 1871-6873},
  doi = {10.1617/s11527-025-02664-3},
  urldate = {2025-10-01},
  langid = {english}
}

@article{tencate1996slow,
  title = {Slow Dynamics in the Nonlinear Elastic Response of {{Berea}} Sandstone},
  author = {TenCate, James A. and Shankland, Thomas J.},
  year = {1996},
  journal = {Geophysical Research Letters},
  volume = {23},
  number = {21},
  issn = {0094-8276, 1944-8007},
  doi = {10.1029/96GL02884},
  urldate = {2025-09-21},
  copyright = {http://onlinelibrary.wiley.com/termsAndConditions\#vor},
  langid = {english}
}

@inproceedings{tencate2000slow,
  title = {Slow Dynamics Experiments in Solids with Nonlinear Mesoscopic Elasticity},
  booktitle = {{{AIP Conference Proceedings}}},
  author = {TenCate, James A.},
  year = {2000},
  volume = {524},
  pages = {303-306},
  publisher = {AIP},
  address = {Gottingen (Germany)},
  issn = {0094243X},
  doi = {10.1063/1.1309228},
  urldate = {2024-04-17},
  langid = {english}
}

@article{tencate2000universal,
  title = {Universal {{Slow Dynamics}} in {{Granular Solids}}},
  author = {TenCate, James A. and Smith, Eric and Guyer, Robert A.},
  year = {2000},
  journal = {Physical Review Letters},
  volume = {85},
  number = {5},
  issn = {0031-9007, 1079-7114},
  doi = {10.1103/PhysRevLett.85.1020},
  urldate = {2024-02-20},
  langid = {english}
}

@incollection{tencate2002nonlinearity,
  title = {Nonlinearity and Slow Dynamics in Rocks: {{Response}} to Changes of Temperature and Humidity},
  booktitle = {Nonlinear Acoustics at the Beginning of the 21st Century},
  author = {TenCate, J.A. and Duran, J and Shankland, {\relax TJ}},
  year = {2002},
  volume = {2},
  pages = {767-770},
  publisher = {Moscow State University, Faculty of Physics},
  address = {Moscow}
}

@article{tencate2004nonlinear,
  title = {Nonlinear and {{Nonequilibrium Dynamics}} in {{Geomaterials}}},
  author = {TenCate, James A. and Pasqualini, Donatella and Habib, Salman and Heitmann, Katrin and Higdon, David and Johnson, Paul A.},
  year = {2004},
  journal = {Physical Review Letters},
  volume = {93},
  number = {6},
  issn = {0031-9007, 1079-7114},
  doi = {10.1103/PhysRevLett.93.065501},
  urldate = {2024-02-20},
  langid = {english}
}

@article{tencate2011slow,
  title = {Slow {{Dynamics}} of {{Earth Materials}}: {{An Experimental Overview}}},
  shorttitle = {Slow {{Dynamics}} of {{Earth Materials}}},
  author = {TenCate, James A.},
  year = {2011},
  journal = {Pure and Applied Geophysics},
  volume = {168},
  number = {12},
  issn = {0033-4553, 1420-9136},
  doi = {10.1007/s00024-011-0268-4},
  urldate = {2024-01-23},
  langid = {english}
}

@article{tremblay2010probing,
  title = {Probing Slow Dynamics of Consolidated Granular Multicomposite Materials by Diffuse Acoustic Wave Spectroscopy},
  author = {Tremblay, Nicolas and Larose, Eric and Rossetto, Vincent},
  year = {2010},
  journal = {The Journal of the Acoustical Society of America},
  volume = {127},
  number = {3},
  issn = {0001-4966, 1520-8524},
  doi = {10.1121/1.3294553},
  urldate = {2024-07-02},
  langid = {english}
}

@article{trtnik2009prediction,
  title = {Prediction of Concrete Strength Using Ultrasonic Pulse Velocity and Artificial Neural Networks},
  author = {Trtnik, Gregor and Kav{\v c}i{\v c}, Franci and Turk, Goran},
  year = {2009},
  journal = {Ultrasonics},
  volume = {49},
  number = {1},
  issn = {0041624X},
  doi = {10.1016/j.ultras.2008.05.001},
  urldate = {2025-10-01},
  langid = {english}
}

@article{vandenabeele2000nonlinear_1,
  title = {Nonlinear {{Elastic Wave Spectroscopy}} ({{NEWS}}) {{Techniques}} to {{Discern Material Damage}}, {{Part I}}: {{Nonlinear Wave Modulation Spectroscopy}} ({{NWMS}})},
  shorttitle = {Nonlinear {{Elastic Wave Spectroscopy}} ({{NEWS}}) {{Techniques}} to {{Discern Material Damage}}, {{Part I}}},
  author = {Van Den Abeele, K. E.-A. and Johnson, P. A. and Sutin, A.},
  year = {2000},
  journal = {Research in Nondestructive Evaluation},
  volume = {12},
  number = {1},
  issn = {0934-9847, 1432-2110},
  doi = {10.1080/09349840009409646},
  urldate = {2025-10-01},
  langid = {english}
}

@article{vandenabeele2000nonlinear_2,
  title = {Nonlinear {{Elastic Wave Spectroscopy}} ({{NEWS}}) {{Techniques}} to {{Discern Material Damage}}, {{Part II}}: {{Single-Mode Nonlinear Resonance Acoustic Spectroscopy}}},
  shorttitle = {Nonlinear {{Elastic Wave Spectroscopy}} ({{NEWS}}) {{Techniques}} to {{Discern Material Damage}}, {{Part II}}},
  author = {Van Den Abeele, K. E.-A. and Carmeliet, J. and Ten Cate, J. A. and Johnson, P. A.},
  year = {2000},
  journal = {Research in Nondestructive Evaluation},
  volume = {12},
  number = {1},
  issn = {0934-9847, 1432-2110},
  doi = {10.1080/09349840009409647},
  urldate = {2025-10-01},
  langid = {english}
}

@article{vandenabeele2002influence,
  title = {Influence of Water Saturation on the Nonlinear Elastic Mesoscopic Response in {{Earth}} Materials and the Implications to the Mechanism of Nonlinearity},
  author = {Van Den Abeele, K. E.-A. and Carmeliet, J. and Johnson, P. A. and Zinszner, B.},
  year = {2002},
  journal = {Journal of Geophysical Research: Solid Earth},
  volume = {107},
  number = {B6},
issn = {0148-0227},
  doi = {10.1029/2001JB000368},
  urldate = {2025-10-01},
  copyright = {http://onlinelibrary.wiley.com/termsAndConditions\#vor},
  langid = {english}
}

@article{yoritomo2020slow_1,
  title = {Slow Dynamics in a Single Glass Bead},
  author = {Yoritomo, John Y. and Weaver, Richard L.},
  year = {2020},
  journal = {Physical Review E},
  volume = {101},
  number = {1},
  issn = {2470-0045, 2470-0053},
  doi = {10.1103/PhysRevE.101.012902},
  urldate = {2024-03-26},
  langid = {english}
}

@article{yoritomo2020slow_2,
  title = {Slow Dynamic Nonlinearity in Unconsolidated Glass Bead Packs},
  author = {Yoritomo, John Y. and Weaver, Richard L.},
  year = {2020},
  journal = {Physical Review E},
  volume = {101},
  number = {1},
  issn = {2470-0045, 2470-0053},
  doi = {10.1103/PhysRevE.101.012901},
  urldate = {2024-04-17},
  langid = {english}
}

@article{yoritomo2020slow_3,
  title = {Slow Dynamic Elastic Recovery in Unconsolidated Metal Structures},
  author = {Yoritomo, John Y. and Weaver, Richard L.},
  year = {2020},
  journal = {Physical Review E},
  volume = {102},
  number = {1},
  issn = {2470-0045, 2470-0053},
  doi = {10.1103/PhysRevE.102.012901},
  urldate = {2024-04-17},
  langid = {english}
}

@article{yoritomo2025slow,
  title = {Slow Dynamic Nonlinear Elasticity during and after Conditioning, a Unified Theory and a Lock-in Probe},
  author = {Yoritomo, John Y. and Weaver, Richard L.},
  year = {2025},
  journal = {Journal of the Mechanics and Physics of Solids},
  volume = {200},
    publisher = {Elsevier BV},
issn = {0022-5096},
  doi = {10.1016/j.jmps.2025.106149},
  urldate = {2025-07-24},
  copyright = {https://www.elsevier.com/tdm/userlicense/1.0/},
  langid = {english}
}

@article{zeng2023temperature,
  title = {Temperature Correction in Acoustoelastic Coefficient Measurements},
  author = {Zeng, Shengyang and Malone, Clayton and Zhu, Jinying},
  year = {2023},
  journal = {NDT \& E International},
  volume = {140},
issn = {09638695},
  doi = {10.1016/j.ndteint.2023.102959},
  urldate = {2024-01-24},
  langid = {english}
}

@article{zeng2026calculation,
  title = {Calculation of Relative Velocity Change of Coherent Waves Using Improved Stretching Technique},
  author = {Zeng, Shengyang and Tu, Jiuzhou and Malone, Clayton and Zhu, Jinying and Li, Xiongbing},
  year = {2026},
  journal = {Ultrasonics},
  volume = {158},
  issn = {0041624X},
  doi = {10.1016/j.ultras.2025.107824},
  urldate = {2025-10-01},
  langid = {english}
}

@article{zhong2022applications,
  title = {Applications of {{Stretching Technique}} and {{Time Window Effects}} on {{Ultrasonic Velocity Monitoring}} in {{Concrete}}},
  author = {Zhong, Bibo and Zhu, Jinying},
  year = {2022},
  journal = {Applied Sciences},
  volume = {12},
  number = {14},
  issn = {2076-3417},
  doi = {10.3390/app12147130},
  urldate = {2025-07-30},
  copyright = {https://creativecommons.org/licenses/by/4.0/},
  langid = {english}
}

@article{alexander2019durability,
  title = {Durability, Service Life Prediction, and Modelling for Reinforced Concrete Structures -- Review and Critique},
  author = {Alexander, Mark and Beushausen, Hans},
  year = {2019},
  journal = {Cement and Concrete Research},
  volume = {122},
  pages = {17--29},
  issn = {00088846},
  doi = {10.1016/j.cemconres.2019.04.018},
  urldate = {2025-10-03},
  langid = {english}
}

@phdthesis{bittner2018understanding,
  author = {Bittner, James Alan},
  title  = {Understanding and Predicting Transient Material Behaviors Associated with Mechanical Resonance in Cementitious Composites},
  school = {University of Illinois at Urbana--Champaign},
  year   = {2018},
  address = {Urbana, IL},
  type   = {Ph.D. thesis},
}

@article{alqurashi2025review,
  title={A Review of Ultrasonic Testing and Evaluation Methods with Applications in Civil NDT/E},
  author={Alqurashi, Inad and Alver, Ninel and Bagci, Ulas and Catbas, Fikret Necati},
  journal={Journal of Nondestructive Evaluation},
  volume={44},
  number={2},
  pages={53},
  year={2025},
  publisher={Springer},
  doi={10.1007/s10921-025-01190-0}
}

@article{asnar2025anisotropy,
  title={Anisotropy reveals contact sliding and aging as a cause of post-seismic velocity changes},
  author={Asnar, Manuel and Sens-Sch{\"o}nfelder, Christoph and Bonnelye, Audrey and Curtis, Andrew and Dresen, Georg and Bohnhoff, Marco},
  journal={Nature Communications},
  volume={16},
  number={1},
  pages={7587},
  year={2025},
  publisher={Nature Publishing Group UK London}
}

@article{simpson2023temperature,
  title={Temperature-Induced Nonlinear Elastic Behavior in Berea Sandstone Explained by a Modified Sheared Contacts Model},
  author={Simpson, Jonathan and van Wijk, Kasper and Adam, Ludmila and Esteban, Lionel},
  journal={Journal of Geophysical Research: Solid Earth},
  volume={128},
  number={1},
  pages={e2022JB025452},
  year={2023},
  publisher={Wiley Online Library}
}

@article{seo2025utilizing,
  title={Utilizing linear and nonlinear ultrasound for improved estimation of concrete strength},
  author={Seo, Wanhyuk and Lee, Seungjun and Baek, Seungo and Shin, Dong Min and Youm, Kwangsoo and Kim, Bongjun and Kim, Gun and Yun, Tae Sup},
  journal={Nondestructive Testing and Evaluation},
  pages={1--17},
  year={2025},
  publisher={Taylor \& Francis}
}

@article{payan2014quantitative,
  title={Quantitative linear and nonlinear resonance inspection techniques and analysis for material characterization: Application to concrete thermal damage},
  author={Payan, Cedric and Ulrich, Timothy J and Le Bas, Pierre-Yves and Saleh, T and Guimaraes, Maria},
  journal={The Journal of the Acoustical Society of America},
  volume={136},
  number={2},
  pages={537--546},
  year={2014},
  publisher={AIP Publishing}
}

@article{shkolnik_2005,
	title = {Effect of nonlinear response of concrete on its elastic modulus and strength},
	volume = {27},
	doi = {10.1016/j.cemconcomp.2004.12.006},
	language = {en},
	number = {7-8},
	urldate = {2019-10-25},
	journal = {Cement and Concrete Composites},
	author = {Shkolnik, I.E.},
	month = aug,
	year = {2005},
	pages = {747--757},
	}
\end{document}